\documentclass[a4paper,11pt]{article}

\usepackage{jheppub} 

\usepackage[T1]{fontenc} 
\usepackage{epstopdf}
\newcommand{\hs}{\hspace*{0.3cm}}

\newcommand{\be}{\begin{equation}}
	\newcommand{\ee}{\end{equation}}
\newcommand{\bea}{\begin{eqnarray}}
	\newcommand{\eea}{\end{eqnarray}}
\newcommand{\ben}{\begin{enumerate}}
	\newcommand{\een}{\end{enumerate}}
\newcommand{\bit}{\begin{itemize}}
	\newcommand{\eit}{\end{itemize}}
\newcommand{\bde}{\begin{widetext}}
	\newcommand{\ede}{\end{widetext}}
\newcommand{\nn}{\nonumber}
\newcommand{\crn}{\nonumber \\}

\newcommand{\al}{\alpha}
\newcommand{\la}{\lambda}

\newcommand{\ga}{\gamma}
\newcommand{\va}{\varphi}

\newcommand{\pa}{\partial}

\newcommand{\fr}{\frac}

\newcommand{\bc}{\begin{center}}
	\newcommand{\ec}{\end{center}}
\newcommand{\Ga}{\Gamma}
\newcommand{\de}{\delta}

\newcommand{\ep}{\epsilon}

\newcommand{\si}{\sigma}

\newcommand{\eq}{\eqref}
\newcommand{\lb}{\label}

\newcommand{\mathsym}[1]{{}}
\newcommand{\gev}{~\mathrm{GeV}}

\def\three{\ensuremath{\mathbf{3}}}
\def\threeS{\ensuremath{\mathbf{3^*}}}

\def\gsim{\raise0.3ex\hbox{$\;>$\kern-0.75em\raise-1.1ex\hbox{$\sim\;$}}}
\def\lsim{\raise0.3ex\hbox{$\;<$\kern-0.75em\raise-1.1ex\hbox{$\sim\;$}}}

\begin{document}
	\title{\boldmath{On axion properties in
		the 3-3-1 model with $U(1)_{B-L}$ and Peccei-Quinn symmetries
	}}
	\author{H.N. Long$^{a,b}$}
	\emailAdd{hoangngoclong@vlu.edu.vn} 
\author{H.T. Hung$^{c,d}$}
\emailAdd{hathanhhung@hpu2.edu.vn}	
\author{V.H. Binh$^{e,d}$}
\emailAdd{vhbinh@iop.vast.vn}	
\author{A.B. Arbuzov$^{d}$}
\emailAdd{arbuzov@theor.jinr.ru}
\affiliation{
	$^a$ Subatomic Physics Research Group,
	Science and Technology Advanced Institute,\\
	Van Lang University, Ho Chi Minh City, Vietnam \\
	$^b$ Faculty of Applied Technology, School of  Technology,  Van Lang University, Ho Chi Minh City, Vietnam \\
	$^c$ Department of Physics, Hanoi Pedagogical University 2, Phuc Yen,  Vinh Phuc 15000, Vietnam\\
	$^d$ Bogoliubov Laboratory of Theoretical Physics, Joint Institute for Nuclear Research, Dubna, 141980 Russia\\
	$^e$ Institute of Physics, Vietnam Academy Science and Technology, 10 Dao Tan, Ba Dinh, Hanoi 10000, Vietnam }

\date{\today }

\abstract{	
	The Peccei-Quinn ($PQ$) mechanism is applied to the $\mathrm{SU(3)_c \otimes SU(3)_L \otimes U(1)_X}$ model with $U(1)_{B-L}$ symmetry. The structures in the $PQ$ charges of all fermions and scalar fields in the model are investigated by applying the invariant condition under the symmetry group transformations on all Yukawa interaction terms. All defined $PQ$ charges depend just on the $PQ$ charge of the complex singlet scalar field which causes the $U(1)_{PQ}$ spontaneous symmetry breaking in the model. The mixing and mass hierarchy in the scalar sector of the model are studied in detail. The constraints on the $PQ$ charges and imaginary parts of scalars are derived. It is shown that only neutral scalar fields carry $PQ$ charges while charged ones do not.
    As the result, the physical state of axion which obeys the invariance under $\mathrm{SU(3)_L \otimes U(1)_X}$ and $PQ$ transformations, is a linear combination of all imaginary parts associated with the $X$ charges of scalar triplets.
    The anomaly axion-fermion interactions are presented.
	Explicit expressions for axion and light Standard Model (SM) like Higgs boson are shown. Mass of the axion and its coupling to photon are derived.
	
	The decays of the SM-like Higgs boson into a pair of either charged leptons
	or bottom quarks are presented and constrained.
	The triple-coupling axion-photon-photon arisen from kinetic terms of scalar fields is derived. Hence, the decay of axion into a pair of photons consists of two parts: the first is related to the anomaly coupling and the second is come from kinetic terms of scalar fields. The result shows that the new contribution can be helpful for searching axion with mass at hundred keV.   }



\maketitle
\noindent

\section{Introduction}
\label{sec1}
An elegant solution of the strong $CP$ problem is the Peccei-Quinn ($PQ$) mechanism
\cite{pq1,pq2}. Spontaneous breaking of the global chiral $U(1)_{PQ}$ symmetry gives rise
to a Goldstone boson called axion~\cite{we1,wi1}. Axion is a popular candidate for dark matter particles. Axion-like particles naturally appear also in many extensions of the Standard Model (SM).
Axion can have a coupling with two photons and one of those can be taken from an {\it external} electromagnetic field. This opens a promising possibility for experimental searches. There are several experiments which look for
axions and try to measure its coupling to photons by means of the {Primakoff effect \cite{agu,s5,s6,s7}} and optical interferometry, e.g.~\cite{ADMX 2018gho,CAST 2017uph} see also Ref.~\cite{Adams 2022pbo} for a recent review.

The 3-3-1 models~\cite{ppf1,ppf2,ppf3,ppf4,flt1,flt2,flt3,flt4,flt5} have several nice features such as explanation of the number of fermion generations and of electric charge quantization. They automatically provide the $PQ$ formalism~\cite{pal}. The latter was implemented in such models two decades ago~\cite{a1,a2,a3,julio1,alp331,jpf}. In these works, the $PQ$ charges  were determined for concrete fields or components of multiplets.
However, in Ref.~\cite{julio2}, the $PQ$ charges are determined for multiplets, i.e., the charges of all members of a given multiplet are equal.
However, as can be seen below, the $PQ$ charges presented in~\cite{julio2}, do not comply with the general rules.
The aim of our article is to study the relations between $PQ$ charges of different multiplets by presenting $PQ$ charged of particles of the model via only one parameter $PQ_\sigma$ which is attributed to the singlet scalar field $\sigma$ that triggers the breaking of the $PQ$ symmetry. 

Furthermore, the relations between the vacuum expectation values (VEVs) and $X$ charges of scalar fields are shown. Especially, by obeying the invariance of the model's gauge symmetry group, the physical state of axion is represented via four imaginary parts of scalars with coefficients including the ratios of $PQ$ charges, VEVs and $X$ charges of scalar fields. This representation is consistent with the expression defined from the mass mixing matrix of $CP$ odd sector of the model. The $CP$ even sector of the model is also reconsidered in order to point out the SM-like Higgs boson and study some kind of SM-like Higgs boson decays. Other scalar sectors such as the charged scalar sector and the complex electrically neutral scalar sector are also considered.

This  paper consists of 8 Sections. Section \ref{sec1} is for introduction.
The particle content, Yukawa interactions, and the Higgs potential are briefly presented in Section~\ref{sec2}. Section~\ref{HiggsSector} is devoted to the Higgs sector where all physical scalar fields including axion and the SM-like Higgs boson are derived. Some decays of the SM-like Higgs boson to a pair of charged leptons and quarks and the corresponding constraints are given in Section~\ref{smh}.
Definition of the $PQ$ charges for multiplets and some relations among their values of scalar triplets
are given in Section~\ref{PQchargeSec}. The constraint equation, the representation of the axion field and the anomaly axion-photon coupling are pointed out in Section~\ref{some}. The triple-coupling of axion-photon-photon arisen from kinetic terms of scalar fields is studied in Section~\ref{Agaga}.
Finally, all main results of this article are summarized in Section~\ref{sec6}.

\section{The model}
\label{sec2}
\subsection{Particle content and $U(1)_{(\mathcal{B}- \mathcal{L})}$}
\lb{sec21}
To overpass the Strong $CP$ Problem with the help of the $PQ$ mechanism and provide neutrino mass through type-I Dirac seesaw mechanism, a vector-like singlet lepton $S_a$ and a singlet scalar $\si$ are added to the particle content of the 3-3-1 model with right-handed neutrinos~\cite{flt1,flt2,flt3,flt4,flt5}.

The generators are defined as in Ref.~\cite{julio2}
\bea
Q & = & T_3 - \fr{1}{\sqrt{3}} T_8 + X~,\crn
B-L & = & -\fr{4}{\sqrt{3}} T_8 + N \,,\label{pt1}
\eea
where $U(1)_{B-L}$ is the residual subgroup after the spontaneous symmetry breaking (SSB).

Note that $N$ defined in  Eq.\eq{pt1} is {\it common} for fermion multiplets~\cite{joshi,cl}
\bea
L & = &  \fr{4}{\sqrt{3}} T_8 + \mathcal{L} I \,,\lb{pt2}\\
B & = & \mathcal{B} I\,,\lb{pt3}
\eea
where $T_3,T_8$ denote for generators of the $SU(3)_L$ group and $I$ is the $3 \times 3$ unit generator.
Then
\bea
B-L & = &  \mathcal{B} I  -\fr{4}{\sqrt{3}} T_8 - \mathcal{L} I =  -\fr{4}{\sqrt{3}} T_8 +
( \mathcal{B} -  \mathcal{L}) \,,\lb{pt2t}\\
\Rightarrow  N  &= &  \mathcal{B} -  \mathcal{L}\,,
\lb{pt4}
\eea
and the values of $\mathcal{B}$ and $\mathcal{L}$ are given in Ref.~\cite{cl}.

In the 3-3-1 model, leptons come in triplets
\bea\label{lep}
\psi_{aL}&=&\left( \nu_{aL}, e_{aL},(\nu_{aR})^{c}\right)^{T} \sim \left(\mathbf{1},\three,-\fr{1}{3},-\fr{1}{3}\right),\\
e_{aR} &\sim& \left(\mathbf{1},\mathbf{1},-1,-1\right),\nn
\eea
where $a=1,2,3$, and the numbers in parentheses represent the field's transformations under the groups $\mathrm{SU(3)_C}$, $\mathrm{SU(3)_L}$, $\mathrm{U(1)_X}$, and $\mathrm{U(1)_N}$, respectively.

One quark family transforms differently from the other two
\bea\label{LHqua}
Q_{\al L}&=&\left( d_{\al L},-u_{\al L},D_{\al L}\right)^{T}\sim\left(\mathbf{3},\threeS,0,-\fr{1}{3}\right),\\
Q_{3L}&=&\left( u_{3L}, d_{3L},U_{3L}\right)^{T}\sim\left(\mathbf{3},\three,\fr{1}{3},1\right),\nn
\eea
with $\al=1,2$.
As usually, right-handed fermions are singlet under the
$\mathrm{SU(3)_L}$ symmetry.
\bea\label{RHqua}
u_{a R}&\sim&\left(\mathbf{3},\mathbf{1},\fr{2}{3},\fr{1}{3}\right),\quad\quad U_{3R}\sim \left(\mathbf{3},\mathbf{1},\fr{2}{3},\fr{7}{3}\right),\nn\\
d_{a R}&\sim& \left(\mathbf{3},\mathbf{1},-\fr{1}{3},\fr{1}{3}\right), \quad\quad D_{\al R}\sim\left(\mathbf{3},\mathbf{1},-\fr{1}{3},-\fr{5}{3}\right).
\eea

To have sufficiently small neutrino masses, vector-like neutral fermions are added
\be
S_{aL,R} \sim  \left(\mathbf{1},\mathbf{1},0,-1\right).
\label{a301}
\ee

In the scalar sector, as usually, there are three fields in the fundamental $\mathrm{SU(3)_L}$ representation
\bea\label{sca}
\Phi_1 =\left(\phi_1^0, \phi_1^-, \tilde{\phi}_1^0\right)^{T}&\sim&\left(\mathbf{1},\three,-\frac{1}{3},\frac{2}{3}\right),\nn\\
\Phi_2=\left(\phi_2^+, \phi_2^0, \tilde{\phi}_2^-\right)^{T}&\sim&\left(\mathbf{1},\three,\frac{2}{3},\frac{2}{3}\right),\\	
\Phi_3=\left(\phi_3^0, \phi_3^-, \tilde{\phi}_3^0\right)^{T}&\sim &\left(\mathbf{1},\three,-\frac{1}{3},-\frac{4}{3}\right)\,, \nn
\eea
and one scalar gauge singlet
\be
\si \sim  \left(\mathbf{1},\mathbf{1},0,0\right).
\label{au302}
\ee
The particle content of the model with quantum numbers is summarized in Table~\ref{qtn}.
\begin{table*}[h]
	\begin{center}
		\begin{tabular}{|l|c|c|c|c|c|c|c|c|c|c|c|c|c|c|}
			\hline
			Multiplet & $\Phi_3$ & $\Phi_1$ & $\Phi_2$ &  $Q_{3L}$ & $Q_{\al L}$ &
			$u_{aR}$&$d_{aR}$ &$U_{3R}$ & $D_{\al R}$ & $f_{aL}$ & $l_{aR}$& $S_L$& $S_R$& $\si$  \\
			\hline $\cal B$ charge &$0$ & $ 0  $ &
			$ 0  $ &  $\fr 1 3  $ & $\fr 1 3  $& $\fr 1 3  $ &
			$\fr 1 3  $ &  $\fr 1 3  $&  $\fr 1 3  $&
			$0  $& $0   $& $0$& $0$& $0$ \\
			\hline $\cal L$ charge &$\fr 4 3$ & $-\fr 2 3  $ &  $-\fr 2 3  $ &
			$-\fr 2 3  $ & $\fr 2 3  $& 0 & 0 & $-2$& $2$&
			$\fr 1 3  $& $ 1   $&$-1$&$-1$&$0$\\
			\hline $N$ charge &$-\fr 4 3$ & $\fr 2 3  $ &  $\fr 2 3  $ &
			$ 1 $ & $- \fr 1 3  $& $\fr 1 3 $ & $\fr 1 3 $ & $\fr 7 3 $& $- \fr 5 3$&
			$- \fr 1 3  $& $ - 1   $&$1$&$1$&$0$\\
			\hline
		\end{tabular}
        \caption{
			${\cal B},{\cal L}$ and  $N $ charges for multiplets in
			the 3-3-1 model with  $U_{\mathcal{B-L}}$ symmetry.} \label{qtn}
	\end{center}
\end{table*}

Using Eq.\eq{pt2t}, one obtains the $\mathcal{B}-\mathcal{L}$ numbers for components of multiplets given in the third column in Table~\ref{tab2}.

Looking at Table \ref{tab2}, one has some inspired remarks that extra quarks $U_3$ and $D_\al$ carry lepton number of two. The situation is  the same for bottom elements of $\Phi_1 $ and $\Phi_2$ and two upper elements of $\Phi_3 $. One more remark only scalar fields with vanishing $\mathcal{B}-\mathcal{L}$ number can have  vacuum expectation values (VEVs).

The masses of  fermions and gauge bosons request VEVs of    scalar  fields as given below
\bea
\Phi_1 = \left( \begin{array}{c}
	\fr{1}{\sqrt2}\left(v_1 +R_1 +i I_1 \right)  \\
	\phi_1^-\\
	\widetilde{\phi}_1^0
\end{array} \right), \hs
\Phi_2 = \left( \begin{array}{c}
	\phi_2^+\\
	\fr{1}{\sqrt2}\left(v_2 +R_2 +i I_2 \right)  \\
	\widetilde{\phi}_2^+
\end{array} \right), \hs
\Phi_3 = \left( \begin{array}{c}
	\phi_3^0 \\
	\phi_3^-\\
	\fr{1}{\sqrt2}\left(v_3 +R_3 +i I_3 \right)
\end{array} \right), \crn\label{VEVtriplet}
\eea
and the neutral singlet scalar field is
\be
\si = \fr{1}{\sqrt2}\left( v_\sigma + R_\sigma + i I_\sigma \right)\,. \label{VEVsinglet}
\ee

\subsection{Yukawa interactions}\lb{sec22}
The Yukawa interactions are given as
\be \mathcal{L}_{\rm Y} = \mathcal{L}_{\rm Y}^l +\mathcal{L}_{\rm Y}^q \,.
\lb{au305}
\ee
The first term on the right-handed side above is devoted to leptons
\bea
-\mathcal{L}_{\rm Y}^l &=&  \,
y^{e}_{ab}\,\overline{\psi_{aL} }\,  \Phi_2 e_{bR}  + y^{\nu_1}_{ab} \,\overline{ \psi_{aL}}\,\Phi_1 \,S_{bR} + y^{\nu_2}_{ab} \,\overline{ \psi_{aL}}\,\Phi_3 \, (S_{bL})^c +
y^{S}_{ab}\, \overline{S_{aL}} \,\sigma^*\, S_{bR}  + \mathrm{H.c.}\crn
\lb{au307}
\eea
So, the charged lepton get masses simply as
\be
M_e = \fr{y^e_{ab} v_2}{\sqrt{2}}\,.
\ee
Again here the mass matrix is diagonalized as $\mbox{diag}(m_e, m_\mu, m_\tau)=(U_{L}^{e})^\dagger M_e U_{R}^{e}$,
where $U_{L,R}^{e}$ are the unitary matrices connecting to the left/right flavor, $e_{L,R}$, and mass eigenstates, $e^\prime_{L,R}$.

Note that the term $\overline{\psi_L} \Phi_2^* (\psi_L)^c$ with an unsuppressed
Dirac neutrino mass, is forbidden due to the $PQ$ symmetry since it is not invariant (see Table \ref{tab2} below)
\[
PQ_{\bar{\psi}_L} + PQ_{\Phi^*} + PQ_{(\psi_L)^c} = \frac{2 PQ_\sigma}{3}\neq 0 \,. \]

\begin{figure}[h!]
	\centering
	\begin{tabular}{c}
		\includegraphics[width=10cm]{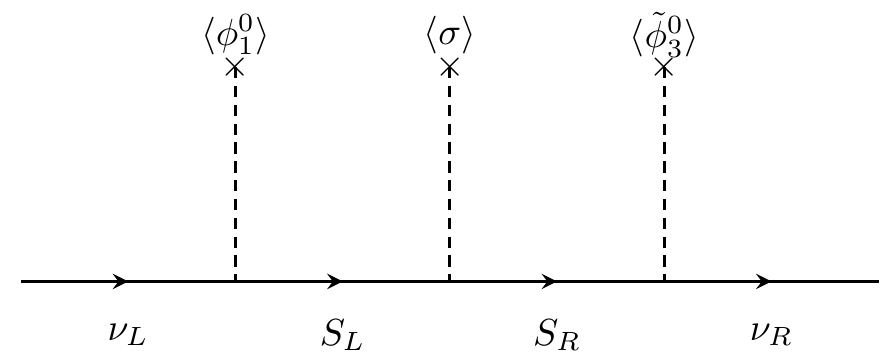}
	\end{tabular}
\caption{ Type I Dirac seesaw mechanism for neutrino mass }
	\label{fig1}
\end{figure}

As a result, neutrino masses are generated via the type-I Dirac seesaw mechanism, illustrated in Fig.~\ref{fig1} (see also in Ref.~\cite{julio2}).

In the basis $N=(\nu, S)$, we can write the neutral lepton mass term $\overline{N_L} M_{\text{Dirac}} N_R$ in terms of the seesaw-type-I matrix,
\bea \label{ssm}
M_{\mathrm{Dirac}} =\fr{1}{\sqrt{2}}\begin{pmatrix}
	0 & y_{ab}^{\nu_1} v_1 \\ (y_{ab}^{\nu_2})^T v_3 & y_{ab}^{S} v_\sigma
\end{pmatrix},
\eea
where ``Dirac'' indicates that all mass terms are of the Dirac type. This matrix can be written in a diagonal form as $\mbox{diag}(m^N_n)=(U_L^N)^\dagger M_D(U_R^N)$, with $n=1,...,6$, once the
chiral flavor fields are rotated to the mass basis through unitary transformations $N_{L,R} \to U_{L,R}^N N_{L,R}^\prime$.
The full Dirac seesaw expansion formula is readily obtained by the method from Ref.~\cite{schechter81}, though here it suffices for us to keep just the first order term,
\be \label{ssm2}
m_\nu^D \simeq \fr{ y_{ab}^{\nu_1} (y_{ab}^S)^{-1} (y_{ab}^{\nu_2})^T }{\sqrt{2}}   \fr{v_1 v_3}{v_\sigma}\,.
\ee
So, one can see how small active neutrino masses result from the suppression by the large seesaw mediator mass, which is identified to lie at the Peccei-Quinn energy scale.
Choosing $v_\si \simeq v_{PQ}$ suggests the existence of new physics at a \textit{lower} scale $v_3$, characterizing the extended electroweak gauge sector of the 3-3-1 model with right-handed neutrino~\cite{flt1}.
For example with $v_1 = 10^2$~GeV, $v_3= 10^{4}$~GeV, and $v_\si = 10^{12}$~GeV, sub-eV neutrino masses ($\sim 0.1$~eV) are obtained for reasonable values of the Yukawa couplings $y^{\nu_{1,2}}\sim 10^{-2}$ and $y^{S}\sim 1$.

Likewise, the lepton mixing matrix describing neutrino oscillations arises as
\be
V_{LEP}=(U_{L}^e)^\dagger U_{L}^\nu\,.
\lb{au316}
\ee

The second term in  Eq.~\eq{au305} is associated with quarks
\bea
-\mathcal{L}_{\rm Y}^q &=& \,
y^{u}_{\al a}\, \overline{Q_L^{\al}}\,\Phi_2^{*} \, u^{a}_R   +
y^{u}_{3a}\, \overline{Q_L^{3}}\, \Phi_1 \,u^{a}_R
+ y^{d}_{3a} \, \overline{Q_L^{3}}\, \Phi_2 \, d^{a}_R  +
y^{d}_{\al a} \, \overline{Q_L^{\al}}\,\Phi_1^{*} \,d^{a}_R \crn 
&+& y^{U}_{33}\, \overline{Q_L^{3}} \,\Phi_3  \,U^3_R
+  y^{D}_{\al\beta} \, \overline{Q_L^{\al}}\,\Phi_3^{*}\,D_R^{\beta}
+ \mathrm{H.c.}
\lb{au306}
\eea
Due to the difference in the $\mathcal{B}-\mathcal{L}$ number of the exotic quark $T$  with  $\fr 7 3$ and the ordinary quarks $u_a$ with  $\fr 1 3$, there is no mixing among them. Hence, in the basis $(u_{a},U_3)$, the following mass mixing appears
\be
M_{u}=\fr{1}{\sqrt{2}}\left(
\begin{array}{cccc}
	-v_2 y^u_{11} & -v_2  y^u_{12} &  -v_2  y^u_{13} & 0 \\
	-v_2 y^u_{21} &  -v_2 y^u_{22} &  -v_2  y^u_{23} & 0 \\
	v_1 y^u_{31} &  v_1  y^u_{32} &  v_1  y^u_{33} & 0 \\
	0 & 0 & 0 & v_3 y^{U}_{33} \\
\end{array}
\right).
\lb{qmassu}
\ee

Similarly, in the basis $(d_{a},D_{\al})$, the mass matrix of the down-type quarks reads
\be
\begin{split}
	M_{d}&=\fr{1}{\sqrt{2}}\left(
	\begin{array}{ccccc}
		v_1 y^d_{11} & v_1  y^d_{12} &  v_1  y^d_{13} & 0 &0\\
		v_1 y^d_{21} &   v_1 y^d_{22} &  v_1  y^d_{23} & 0 &0\\
		v_2  y^d_{31} &   v_2  y^d_{32} &  v_2  y^d_{33} & 0 &0\\
		0 & 0 & 0 &  v_3 y_{11}^{D} & v_3 y_{12}^{D} \\
		0 & 0 & 0 &  v_3 y_{12}^{D} &  v_3 y_{22}^{D} \\
	\end{array}
	\right).
\end{split}
\lb{qmassd}
\ee
As in the previous case, the mass matrix of the standard down-type quarks, $m^d_{3\times 3}$, is diagonalized by rotating the flavor states to the mass basis
$d_{L,R}\to U^d_{L,R}\, d^\prime_{L,R}$, so as to obtain $ \mbox{diag}(m_d, m_s, m_b)=(U_{L}^{d})^\dagger m_{3\times 3}^d U_{R}^{d}$.

As a consequence, the Cabibbo-Kobayashi-Maskawa matrix describing light quark mixing, defined as
\be
V_{CKM}=(U_{L}^u)^\dagger U_{L}^d,
\lb{au313}
\ee
is the same as in the SM.

\subsection{The Higgs potential}
\lb{sec23}

The Higgs potential in the model reads
\bea
V_H &=& \sum_{i=1}^3 \mu_i^2 \Phi_i^\dagger \Phi_i+ \lambda_i (\Phi^\dagger_i \Phi_i)^2 + \sum_{i,j=1;i<j}^3 \lambda_{ij}(\Phi^\dagger_i \Phi_i)(\Phi^\dagger_j \Phi_j) + \tilde{\lambda}_{ij}(\Phi^\dagger_i \Phi_j)(\Phi^\dagger_j \Phi_i)\crn
&& +\mu_\sigma^2 \sigma^{\ast} \sigma + \lambda_\sigma (\sigma^{\ast} \sigma)^2 + \sum_{i=1}^3 \lambda_{i\sigma}(\Phi^\dagger_i \Phi_i)(\sigma^{\ast} \sigma) + (\lambda_A \,\sigma\, \Phi_1 \Phi_2 \Phi_3 + H.c.). \label{VH}
\eea
The last term in Eq.~\eq{VH} appears with the plus sign and it is typical for the models with axions. We will turn back to this point in Sec.~\ref{PQchargeSec} to define a constraint on $PQ$ charges of scalar fields and to analyze it in detail in Sec.~\ref{HiggsSector}.

\subsection{ Gauge bosons}
\lb{sec24}
There are nine electroweak gauge bosons: eight $W_\mu^i, (i=1,...,8) \in  SU(3)_L$ and one $B_\mu \in U(1)_X$. The kinetic terms of scalar fields are
\be \mathcal{L}_{Higgs} = \sum_{H=\Phi_1,\Phi_2,\Phi_3, \si} (D^\mu H)^\dag D_\mu H \,.
\label{kineticGB}
\ee
The mass eigenstates of charged gauge bosons are
\bea W^{\pm}_{\mu } &= &\fr{1}{\sqrt{2}}\left( W^1_{\mu}\mp i W^2_{\mu}\right),\hs
Y^{\pm }_{\mu} = \fr{1}{\sqrt{2}}\left( W^6_{\mu}\pm i W^7_{\mu}\right),\crn
X^{ 0}_{\mu} & = & \fr{1}{\sqrt{2}}\left( W^4_{\mu}- i W^5_{\mu}\right),\hs
X^{0*}_{\mu}=\fr{1}{\sqrt{2}}\left( W^4_{\mu}+i W^5_{\mu}\right)
\label{j234}\eea
with the corresponding masses
\be m^2_{W} = \fr{g^2}{4}(v_1^2+v_2^2),\hs m^2_{X^0}=\fr{g^2}{4}(v_3^2+v_1^2),\hs m^2_Y=\fr{g^2}{4}(v_{3}^2+v_2^2)\,,\label{mWXY} \ee
where $g$ is the $SU(3)_L$ coupling constant.

Three neutral gauge bosons include photon $A$ and two neutral heavy vector bosons $Z$ and $Z^\prime$, whose masses are respectively defined as
\bea
m^2_A &=& 0\,, \notag\\
m^2_Z &=& \frac{g^2(9g^2 + 2 g_X^2)(v_1^2 v_2^2 + v_2^2 v_3^2 + v_3^2 v_1^2)}{2\left(18g^2(v_1^2 + v_2^2 +v_3^2)+g_X^2 (v_1^2 + 4v_2^2 +v_3^2)\right)}\,, \lb{mAZZp}\\
m^2_{Z^\prime}
&=& \fr{g^2}{3} (v_1^2 + v_2^2 +v_3^2) + \fr{g_X^2}{54}(v_1^2 + 4v_2^2 +v_3^2) - m^2_Z\,,\nn \eea
where $g_X$ is coupling constant of the $U(1)_X$ subgroup.  The mass of $Z^\prime$ boson is almost fully generated from the terms which contain $v_3^2$.  According to Ref.~\cite{padax}, $v_3 \geq 10^4$~GeV for $m_{Z^\prime} \geq 4.1$~TeV.

Note that exotic quarks $U$ and $D_\al$ as well as gauge bosons $X^0$ and $Y^\pm$ carry the lepton number of two~\cite{joshi,cl,jh18}.
The gauge boson couplings in our model are the same as in Refs.~\cite{self1,self2}.
Due to the specific structure of the quark family, there are flavor changing neutral currents mediated by $Z^\prime$ boson at the tree level because of its interaction in the down type quark sector~\cite{fc1,fc2,fc3,lv}.

\section{The Higgs sector} \label{HiggsSector}

With the field content and symmetry properties are shown in Table \ref{tab2}, the scalar potential is given by Eq.(\ref{VH}) as shown in subsection~\ref{sec23}. Note that the neutral triplet scalar fields with different quantum numbers are expanded around their VEVs as in Eq.~(\ref{VEVtriplet}) and the complex singlet is also expanded around its VEV as in Eq.~(\ref{VEVsinglet}).
With the VEVs that satisfy the hierarchy $v_\sigma^2 \gg v_3^2 \gg v_1^2 + v_2^2 = v_{EW}^2$, the extremum conditions at the tree-level are solved simultaneously for the mass parameters $\mu_1, \mu_2, \mu_3$, and $\mu_\sigma$
\bea
&&\mu_\sigma^2 + \lambda_\sigma  v_\sigma^2 + \fr{\lambda_{1 \sigma}}{2}  v_1^2 +\fr{\lambda_{2 \sigma}}{2}
v_2^2+\fr{\lambda_{3 \sigma}}{2}  v_3^2 + \fr{\lambda_A }{2  } \fr{v_1 v_2 v_3}{ v_\sigma} = 0\,, \notag\\
&&   \mu_3^2 + \lambda_3 v_3^2+ \fr{\lambda_{13}}{2}v_1^2 + \fr{\lambda_{23}}{2}v_2^2+ \fr{\lambda_{3\sigma}}{2}v_\sigma^2+\fr{\lambda_A}{2}\fr{v_1 v_2 v_\sigma}{v_3} = 0\,, \notag\\
&&   \mu_2^2 +\lambda_2 v_2^2+ \fr{\lambda_{12}}{2}v_1^2 + \fr{\lambda_{23}}{2}v_3^2+ \fr{\lambda_{2\sigma}}{2}v_\sigma^2+\fr{\lambda_A}{2}\fr{v_1 v_3 v_\sigma}{v_2} = 0\,, \notag\\
&&   \mu_1^2 + \lambda_1 v_1^2+ \fr{\lambda_{12}}{2}v_2^2 + \fr{\lambda_{13}}{2}v_3^2+ \fr{\lambda_{1\sigma}}{2}v_\sigma^2+\fr{\lambda_A}{2}\fr{v_2 v_3 v_\sigma}{v_1} =0. \label{minimalcondition}
\eea
In order to give masses to all fermions and gauge bosons in Beyond the Standard Models (BSMs), SSBs are necessary.
After SSBs by non-zero VEVs, the Higgs potential gains $CP$ even, $CP$ odd, and complex neutral as well as charged scalar mass mixing matrices.

\subsection{CP-odd sector}\label{CPoddSec}
In the basis of $(I_1, I_2, I_3, I_\sigma)$, the squared mass mixing matrix reads
\bea
M_I^2 = -\fr{A}{2} \left(
\begin{array}{cccc}
	\fr{1}{v_1^2} & \fr{1}{v_1 v_2}& \fr{1}{v_1 v_3} & \fr{1}{v_1 v_\sigma} \\
	& \fr{1}{v_2^2}& \fr{1}{v_2 v_3}& \fr{1}{v_2 v_\sigma} \\
	& & \fr{1}{v_3^2} & \fr{1}{v_3 v_\sigma} \\
	& & & \fr{1}{v_\sigma^2}
\end{array}\right), \label{modd}
\eea
where $A = \lambda_A v_1 v_2 v_3 v_\sigma$. The mass mixing matrix in Eq.~(\ref{modd}) is with the minus sign so that, with the assumptions $v_1, v_2, v_3, v_\sigma >0$ and the positivity condition for the squared mass matrix, one concludes that the value of $\lambda_A$ should be negative.

The matrix $M_I^2$ in Eq.(\ref{modd}) is diagonalized by three mixing angles that are given as follows~\cite{alp331}
\bea
\tan \al =  \fr{v_1}{v_2}\,,\hs \tan \al_1 
= \fr{v_1}{v_3}|\cos \al|\,,\hs
\tan \al_2 = \fr{v_3}{v_\sigma} |\sin \alpha_1| 
\,.\label{oddangles}
\eea

From Eq.~(\ref{oddangles}), we see that only angle $\al$ is significant because both of $v_1$ and $v_2$ are at the EW scale. The mixing angles $\al_1, \al_2$ are tiny due to the hierarchy of the scalars' VEVs.

After being diagonalized, the physical states of the CP-odd sector can be defined as
	\bea
	\left(
	\begin{array}{c}
		a\\
		G_{Z^\prime} \\
		G_Z	\\
		A_5	 \\
	\end{array}
	\right)  =
	\left(
	\begin{array}{cccc}
		\cos \al_2 & -\sin \al_1 \sin \al_2 &- \sin
		\al \cos \al_1 \sin \al_2 & -\cos \al
		\cos \al_1 \sin \al_2 \\
		0 & \cos \al_1 & - \sin \al \sin \al_1 & -\cos
		\al \sin \al_1 \\
		0 & 0 & \cos \al & - \sin \al \\
		\sin \al_2 & \sin \al_1 \cos \al_2 &  \sin
		\al \cos \al_1 \cos \al_2 & \cos \al
		\cos \al_1 \cos \al_2 \\
	\end{array}
	\right)
	\left(
	\begin{array}{c}
		I_\sigma\\
		I_{3} \\
		I_2	\\
		I_1 \\
	\end{array}\right).\notag\\
	\label{physodd}
	\eea
Among these four CP-odd scalar physical states, just the $A_5$ field with four components, is massive. Its mass is given by
\begin{eqnarray}
	m_{A_5}^2 = -\fr{A}{2} \left( \sum_{i=1}^3 \fr{1}{v_i^2}+\fr{1}{v_\sigma^2}\right). \label{mA5}
\end{eqnarray}
The above equation for the mass of $A_5$ field is the same as Eq.~(18) in Ref.~\cite{julio2} but the physical state of $A_5$ is determined in a different way. The mass of this $A_5$ field should be above the lower bound of 250~GeV coming from the numerical analysis of the $h \to \mu \mu$ and $h \to \tau \tau$ decays as shown in Ref.~\cite{alp331}.

Three other CP-odd scalar physical states are massless. Two of them are $G_Z$ and $G_{Z^\prime}$ which are Goldstone bosons eaten by $Z$ and $Z^\prime$ bosons, respectively. These Goldstone bosons do not harbor the component $I_\sigma$ which
belongs to the $\sigma$ field. The last massless physical state is the combination of four components $I_1, I_2, I_3, I_\sigma$ which means that this field couples with either ordinary quarks or exotic quarks. This massless field is assumed to be the axion associated with the spontaneous symmetry breaking of the anomaly $U(1)_{PQ}$ and it is expressed as
\be
a = \left( \cos \al_2 I_\sigma - \sin \al_1 \sin \al_2 I_3
- \sin \al \cos \al_1 \sin \al_2 I_2 - \cos \al  \cos \al_1 \sin\al_2 I_1 \right)\,.
\label{axion1}
\ee
The above expression for the axion field, presented via the mixing angles, have never been shown before.
In the limit $v_\sigma\gg v_3 \gg v_1,v_2 \equiv v_{EW}$, the last three terms in the expression for axion in Eq.~(\ref{axion1}) can be omitted. The leading contribution to the axion is the imaginary part of the $\sigma$ field ($I_\sigma$). Hence, the axion is approximately defined by $a \simeq \cos \al_2 I_\sigma$ and it decouples from both the ordinary and exotic quarks in Yukawa interactions.

\subsection{CP-even sector}\label{CPevensector}
In the basis of ($R_1, R_2, R_3, R_\sigma$), the relevant squared mass mixing matrix has the form
\bea
M_R^2 = \left( \begin{array}{cccc}
	2\lambda_1 v_1^2 - \fr{A}{2v_1^2}  & \lambda_{12} v_1 v_2 + \fr{A}{2 v_1 v_2}& \lambda_{23} v_1 v_3 + \fr{A}{2 v_1 v_3} & \lambda_{1\sigma} v_1 v_\sigma +\fr{A}{2 v_1 v_\sigma} \\
	& 2 \lambda_2 v_2^2 - \fr{A}{2v_2^2} &\lambda_{23} v_2 v_3 + \fr{A}{2 v_2 v_3}  & \lambda_{2\sigma} v_2 v_\sigma +\fr{A}{2 v_2 v_\sigma} \\
	&  & 2 \lambda_3 v_3^2 - \fr{A}{2 v_3^2} & \lambda_{3\sigma} v_3 v_\sigma +\fr{A}{2 v_3 v_\sigma} \\
	&  &  & 2 \lambda_\sigma v_\sigma^2 - \fr{A}{2 v_\sigma^2} \\
\end{array}
\right). \label{meven}
\eea
The above matrix is diagonalized within the Hatree-Fock approximation that results in four non-zero eigenvalues with three tiny mixing angles given by
\bea
\tan 2 \beta_1 &=& \fr{4 \cos \beta_2 v_1 v_2 (A + \lambda_{12}v_1^2 v_2^2)}{A \cos^2 \beta_2 v_1^2 - A v_2^2 + 4 v_1^2 v_2^2 (\lambda_1 v_1^2 - \lambda_2 \cos^2 \beta_2 v_2^2)}\,,\crn
\tan 2 \beta_2 &=& \fr{4 v_3 (A + 2 \lambda_{23}v_2^2 v_3^2)}{\cos \al_2 (A - 4\lambda_3 v_3^4)}\,,\hs
\tan 2\beta_3 = \fr{\lambda_{3\sigma}v_3}{\lambda_{\sigma}v_\sigma}\,.\label{evenangles}
\eea
The four physical states which are associated with four massive CP-even scalar bosons $h_5,h,H$, and $\Phi$ are defined as follows~\cite{alp331}
\bea
\left(
\begin{array}{c}
	h_5\\
	h \\
	H	\\
	\Phi \\
\end{array}
\right)  =
\left(
\begin{array}{cccc}
	\cos \beta_1 & \sin \beta_1 \cos \beta_2 & \sin
	\beta_1  \sin \beta_2 \cos \beta_3 & \sin
	\beta_1  \sin \beta_2 \sin \beta_3 \\
	- \sin \beta_1 & \cos \beta_1 \cos \beta_2 & \cos
	\beta_1 \sin \beta_2 \cos \beta_3 & -\cos \beta_1
	\sin  \beta_2 \sin \beta_3 \\
	0 & -\sin \beta_2 & \cos \beta_2 \cos \beta_3 & -\cos
	\beta_2 \sin \beta_3 \\
	0 & 0 & \sin \beta_3 & \cos \beta_3 \\
\end{array}
\right)
\left(
\begin{array}{c}
	R_1\\
	R_2 \\
	R_3	\\
	R_\sigma \\
\end{array}\right)\,,
\label{physeven}
\eea
with their masses given by~\cite{alp331}
	\bea
	m_\Phi^2 &=& v_\sigma \left( \sqrt{\lambda_{3\sigma}^2 v_3^2 + \lambda_\sigma^2 v_\sigma^2}+\lambda_\sigma v_\sigma\right) \approx 2 \lambda_\sigma v_\sigma^2\,,\label{mPhi}\eea
	\bea m_H^2 &=& \lambda_3 v_3^2 + v_3 \sqrt{\lambda_{23}^2 v_2^2 + \lambda_3^2 v_3^2} \approx 2 \lambda_3^2 v_3^2 + \fr{\lambda_{23}^2}{2\lambda_3}v_2^2\,,\label{mH}
    \eea
{\small
\bea	m_{h,h_{5}}^2 &=& \lambda_1 v_1^2 + \lambda_2 v_2^2 - \fr{A v_{EW}^2}{4v_1^2 v_2^2}\notag\\
	&& \pm  \left[ \left( \lambda_1 v_1^2 -\lambda_2 v_2^2\right)^2 -\lambda_{12}^2 v_1^2 v_2^2 - A (\lambda_1+\lambda_2 - \lambda_{12})+ \frac{\lambda_\sigma^2 v_\sigma^2 v_3^2 v_{EW}^2 + 6A\left( \lambda_1 v_1^4 +\lambda_2 v_2^4\right)}{12 v_1^2 v_2^2}  \right]^{\frac{1}{2}}.\crn\label{mhh5}
	\eea
}
The mass of the $\Phi$ field comes almost only from $v_\sigma$ as shown in Eq.~(\ref{mPhi}) while the mass of the $H$ field gets the leading contribution from $v_3$ as follows from Eq.~(\ref{mH}).
For $v_\sigma = 10^{12}$~GeV, the heaviest state is $\Phi \approx \sin \beta_3 R_3 + \cos \beta_3 R_\sigma \simeq \cos \beta_3 R_\sigma$. This state can be used to explain inflation in the Early Universe. With $v_3 = 10^5$ GeV, the $H \approx - \sin \beta_2 R_2 + \cos \beta_2 \cos \beta_3 R_3 - \cos \beta_2 \sin \beta_3 R_\sigma \simeq \cos \beta_2 \cos \beta_3 R_3$ is the new heavy CP-even scalar boson which gets a mass at the TeV scale. The lighter states are $h$ and $h_5$ which are the combinations of 4 components of $R_1, R_2, R_3$, and $R_\sigma$. For the chosen hierarchy of the VEVs of these four scalars, masses of $h$ and $h_5$ come mainly from $v_1$ and $v_2$ so that their masses should be at the EW scale. In summary, the CP-even sector has one heaviest boson with mass at the $10^{12}$~GeV scale, a new heavy boson whose mass is at the TeV scale, and two light bosons with masses at the EW scale.

From the interaction terms in the Higgs potential given in  Eq.~(\ref{VH}), the following reasonable assumption can be proposed: $\lambda_1 \approx \lambda_2 \approx \lambda_{12}$. This allows to simplify the mass equations for $h$ and $h_5$:
\be
m_{h,h_5}^2 \approx \lambda_2 v_{EW}^2 + \fr{m_{A_5}^2}{2} \pm \sqrt{m_{A_5}^4
	+ \lambda_2^2 (v_{EW}^4 - 3v_1^2 v_2^2) - \lambda_2 m_{A_5}^2 \left( v_{EW}^2
	-\fr{2v_1^2 v_2^2}{v_{EW}^2}\right)}.
\ee
In order to make it simpler, we can assume also that $v_1 = v_2 = \fr{v_{EW}}{\sqrt2}$. In this case, the model predicts
\bea
m_h^2 &\simeq& \fr{3}{2} \lambda_2 v_{EW}^2  \,,\\
m_{h_5}^2 &\simeq& \fr{1}{2} \lambda_2 v_{EW}^2 + m_{A_5}^2\,.
\eea
The CP-even scalar $h$ is the SM-like Higgs boson whose mass is of 125~GeV and the new light CP-even scalar $h_5$ gets mass also at the EW scale. The mass of $h_5$ might be of 96~GeV \cite{951,952} or 150~GeV~\cite{1101,1102} and certainly below the TeV scale, i.e., below the scale of $v_3$ where we have the heavier scalar $H$.

\subsection{Complex neutral scalar sector}
In the basis of two components which have opposite $\mathcal{B}-\mathcal{L}$ charges $( \phi_3^{0},\widetilde{\phi}_1^0)$, the squared mass mixing matrix is
\bea
M_{CN}^2= \left(\begin{array}{cc} \widetilde{\lambda}_{13}v_1^2 -\fr{A}{v_3^2}  & \widetilde{\lambda}_{13}v_1 v_3 - \fr{A}{v_1 v_3}\\
	\widetilde{\lambda}_{13}v_1 v_3 - \fr{A}{v_1 v_3}  & \fr{1}{2}
	\widetilde{\lambda}_{13}v_3^2 -\fr{A}{v_1^2}
\end{array} \right). \label{mcomplex}
\eea
As the result, the physical states of complex scalar fields are defined as
\bea
\left( \begin{array}{c}
	G_{X^0} \\
	\varphi^0
\end{array}\right) = \left( \begin{array}{cc}
	\cos \delta & \sin \delta \\
	-\sin \delta & \cos \delta
\end{array}\right)
\left( \begin{array}{c}
	\widetilde{\phi}_1^0 \\
	\phi_3^{0*}
\end{array}\right), \label{complexstates}
\eea
where the tiny mixing angle is given by
\be
\tan \delta = - \fr{v_1}{v_3}\,.
\ee
Eq.~(\ref{complexstates}) shows that there are two physical states in the complex neutral scalar sector. One massless state is $G_{X^0}$, it is associated with a longitudinal component of the neutral gauge boson $X^0$. The other is the massive state $\varphi^0 = \widetilde{\Phi}_1^0\cos \delta  - \Phi_3^{0*}\sin \delta \simeq \widetilde{\Phi}_1^0\cos \delta$ because of the assumption that $v_3 \gg v_1$ which makes $\sin \delta$ to be tiny and possible to be neglected. This state gets mass as 
\be
m^2_{\varphi^0} = \left( \widetilde{\lambda}_{13} -\fr{A}{v_1^2 v_3^2}\right) \fr{ v_1^2 + v_3^2}{2}\,.\label{masscomplex}
\ee
The above equation 
shows that the mass of this neutral complex scalar should be at the energy scale of $v_3$ which is at the TeV scale.

\subsection{Charged scalar sector}
In the basis $(\phi_2^\pm, \phi_1^\pm, \phi_3^\pm, \widetilde{\phi}_2^\pm)$, the squared mass mixing matrix is
\bea
M_\pm^2 = \fr{1}{2}\left( \begin{array}{cccc}
	\widetilde{\lambda}_{12} v_1^2  -\fr{A}{v_2^2}&  \widetilde{\lambda}_{12} v_1 v_2 -\fr{A}{v_1 v_2}& 0 & 0 \\
	\widetilde{\lambda}_{12} v_1 v_2 -\fr{A}{v_1 v_2} & \widetilde{\lambda}_{12} v_2^2  -\fr{A}{v_1^2}  & 0 & 0\\
	0 & 0 & \widetilde{\lambda}_{23} v_2^2  -\fr{A}{v_3^2}  & \widetilde{\lambda}_{23} v_2 v_3 -\fr{A}{v_2 v_3} \\
	0 & 0 & \widetilde{\lambda}_{23} v_2 v_3 -\fr{A}{v_2 v_3} & \widetilde{\lambda}_{23} v_3^2  -\fr{A}{v_2^2}
\end{array}
\right).\label{mcharged}
\eea
The matrix $M_\pm^2$ in Eq.~(\ref{mcharged}) shows that the charged fields with different $\mathcal{B}-\mathcal{L}$ charges do not mix. After being diagonalized, the matrix $M_\pm^2$ gives two massless states $G_{W^\pm} = \Phi_2^\pm \cos \zeta_1  + \phi_1^\pm \sin \zeta_1 $ and $G_{Y^\pm} = \Phi_3^\pm \cos \zeta_2  + \widetilde{\phi}_2^\pm \sin \zeta_2 $ which are associated with the Goldstone bosons eaten by the longitudinal components of $W^\pm$ and $Y^\pm$ charged gauge bosons. We also find two massive physical states
\bea
H_1^\pm = -\phi_2^\pm \sin \zeta_1 + \phi_1^\pm \cos \zeta_1  \,,\hs
H_2^\pm = - \phi_3^\pm \sin \zeta_2  + \widetilde{\phi}_2^\pm \cos \zeta_2 \,,
\eea
with the mixing angles given by
\be
\tan \zeta_1 = - \fr{v_1}{v_2}\,, \hs \tan \zeta_2 = - \fr{v_2}{v_3}\,,
\ee
and the masses of $H_1^\pm $ and $H_2^\pm $ are defined as
\bea
m_{H_1^\pm }^2 = \fr{v_{EW}^2}{2} \left( \widetilde{\lambda}_{12}-\fr{A}{v_1^2 v_2^2}\right) \,, \hs  m_{H_2^\pm }^2 = \fr{v_{2}^2 +v_3^2}{2} \left( \widetilde{\lambda}_{23}-\fr{A}{v_2^2 v_3^2}\right)\,. \label{masschargeHiggs}
\eea
Eq.~(\ref{masschargeHiggs}) shows that the mass of the $H_1^\pm$ field should be at the EW scale while the $H_2^\pm$ field would get a heavier mass at the TeV scale, i.e., {at the scale of $v_3$. }

In the limit $v_\si \gg v_3 \gg v_1 , v_2 $, the scalar content of the model can be presented as follows (in the same way as in the Ref.~\cite{alp331}):
\bea
\Phi_1  \simeq  \left(
\begin{array}{c}
	\fr{1}{\sqrt{2}}\left(v_1 +h_5 + i A_5\right) \\
	H_1^- \\
	G_{X^0}   \\
\end{array}
\right),\, \, \,
\Phi_2 & \simeq &
\left(
\begin{array}{c}
	G_{W^+} \\
	\fr{1}{\sqrt{2}}\left(  v_2 + h+ i G_Z\right)\\
	H_2^+ \\
\end{array}
\right)\,,\,	
\phi_3  \simeq
\left(
\begin{array}{c}
	\chi_1^0 \\
	G_{Y^-} \\
	\fr 1{\sqrt{2}}\left( v_3  + H_\chi + i G_{Z'}\right) \\
\end{array}
\right)\, ,\,\,
\crn
\si & = & \fr{1}{\sqrt{2}} \left( v_\si + \Phi + i a \right)\,.
\label{au281}\eea

\section{Some processes related to the SM-like Higgs boson $h$ \lb{smh}}
	In this section, we consider processes which are possible to be studied theoretically only when SM-like Higgs boson is defined as in subsection \ref{CPevensector},
	such as the decays of the SM-like Higgs boson into a pair of charged leptons as well as its decay into a pair of bottom quarks. These processes have not yet been studied numerically within the model under consideration. 

\subsection{SM-like Higgs boson decay into two down-type quarks $h \to \bar{b}b$}
\begin{figure}[ht]
	\centering
	\begin{tabular}{c}
		\includegraphics[width=12cm]{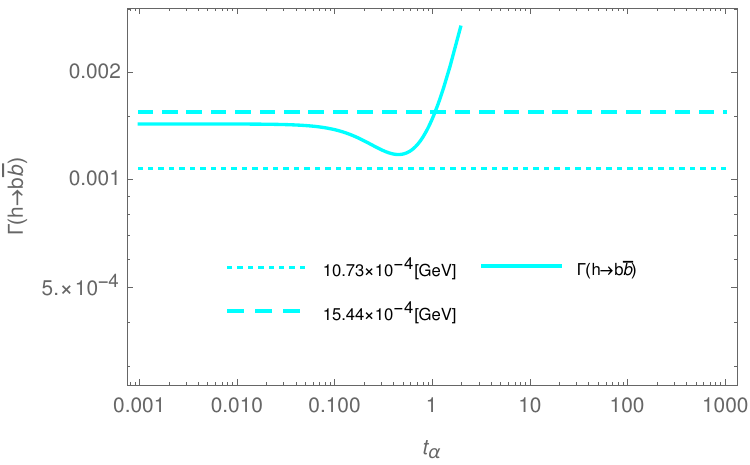}
	\end{tabular}%
	\caption{ Plots of $\Ga (h \rightarrow \bar{\mu}\, \mu)$ and $\Ga (h \rightarrow \bar{\tau}\, \tau)$ depend on $t_{\alpha}$. The dashed lines represent the upper and lower bounds from the experimental data.}
	\label{fig_hbb}
\end{figure}

The decay rate of the process $h \rightarrow \bar{b}b$ is
\bea
\Ga (h \rightarrow \bar{b}b) =  \int d\Ga = \fr{g_{hbb}^2}{8\pi}m_h\left(1-\fr{4m_b^2}{m_h^2}\right)^{\fr 3{2}},
\eea
with \cite{alp331}
\bea
g_{hb\bar{b}}&=&\fr{c_{ \beta _{1}}}{v_2}%
 \left( V_{L}^{\left( d\right) }\right)
^{\dagger } _{33}\left( V_{dL}\widetilde{M}_{d}V_{dR}^{\dagger
}\right) _{33}\left( V_{R}^{\left( d\right) }\right) _{33} -\frac{s_{ \beta _{1}}}{v_1}  \left( V_{L}^{\left( d\right) }\right) ^{\dagger } _{33}\left( V_{dL} \widetilde{M}_{d}V_{dR}^{\dagger }\right) _{33}\left( V_{R}^{\left( d\right) }\right) _{33}\nn\\
&=&  \left(\fr{c_{ \beta_1}  }{v_2}-\fr{\beta _{1}  }{v_1}\right)m_b=\left(\fr{c_{ \beta _{1}}  }{c_{\al}}-\fr{s_{ \beta _{1} } }{s_\al}\right)\fr{m_b}{v}=\frac{c_{\beta_1}}{s_\al}(t_{ \al} -t_{ \beta _{1}})\fr{m_b}{v}= \frac{2 c_{ \beta _{1}}}{c_\al}\frac{m_b}{v}\nn\\
&=&a_{h\bar{b}b}g_{hb\bar{b}}^{SM}.
\eea
where $a_{hb\bar{b}}$ is the deviation factor from the SM Higgs coupling to bottom quarks (in the SM this factor is unity) and $c_\xi=\cos \xi, s_\xi =\sin \xi, t_\xi = \tan \xi$ with $\xi = \beta_1, \beta_2, \alpha$. The experimental data constraint on the $a_{h\bar{b}b}$ parameter is given by \cite{1101,1102}:
\bea \label{limqq}
a_{hb\bar{b}}^{exp}=0.91^{+0.17}_{-0.16} \,,
\eea
with $m_h = 125.09$ GeV,\, $m_{b}=4.183$ GeV, and the decay width limited by CMS as in Eq.~\eq{limqq}, we can give the following bounds on the decay widths of the SM-like Higgs boson into two bottom quarks:
\bea \label{limit_hbb}
10.73  \times 10^{-4} \leq \Ga (h \rightarrow \bar{b}\, b) \leq 15.44 \times 10^{-4}\, \mathrm{GeV}
\eea
In order to numerically investigate $\Ga (h \rightarrow \bar{b}\,b)$,the quantity $t_{\beta_{1,2,3}}$ are parameterized as
\bea
t_{\beta_1} &=& \frac{\lambda_\sigma v_3 v_\sigma \left(v_2^2 -  c_{\beta_2}^2 v_1^2 \right)+4\lambda_2 v_1 v_2 \left( c_{\beta_2}^2 v_2^2 -v_1^2 \right)}{4 c_{\beta_2} v_1 v_2\left(\lambda_{12} v_1 v_2 + \lambda_\sigma v_3 v_\sigma \right)} \notag\\ 
&& - \left[2+ \left( \frac{\lambda_\sigma v_3 v_\sigma \left(v_2^2 -  c_{\beta_2}^2 v_1^2 \right)-4\lambda_2 v_1 v_2 \left( c_{\beta_2}^2 v_2^2 -v_1^2 \right)}{4 c_{\beta_2} v_1 v_2\left(\lambda_{12} v_1 v_2 + \lambda_\sigma v_3 v_\sigma \right)}\right)^2 \right]^{\frac{1}{2}}\,,\\
t_{\beta_2} &=& \left[1 + \frac{c_{\beta_3} (2\lambda_3 v_3^4 +m^2_{A_5}c_\alpha s_\alpha v_1 v_2)}{8 v_2 v_\sigma (\lambda_{23}v_2 v_3^2 - m_{A_5}^2 c_\alpha s_\alpha v_1)} \right]^2 + \frac{c_{\beta_3}^2 (2\lambda_3 v_3^4 +m^2_{A_5}c_\alpha s_\alpha v_1 v_2)^2}{64 v_2^2 v_\sigma^2 (\lambda_{23}v_2 v_3^2 - m_{A_5}^2 c_\alpha s_\alpha v_1)^2}\,,\\
t_{\beta_3}&&=\frac{\sqrt{\lambda _{3\sigma}^2 v_{3 }^2+\lambda _{\sigma}^2 v_{\sigma }^2}-\lambda _{\sigma} v_{\sigma}}{\lambda _{3\sigma} v_{3 }}, \label{tmixangel}
\eea
where the VEVs are expressed through other parameters such as: the mass of the new charged gauge boson ($m_Y^\pm$), the mixing angle $\alpha$ ($t_\alpha=\frac{v_1}{v_2}$), namely,  $v_{3 }=\sqrt{\frac{4m_Y^2}{g^2}-v_2^2},\, v_{1 }=\frac{v t_a}{\sqrt{t_a^2+1}},\, v_{2 }=\frac{v}{\sqrt{t_a^2+1}}$. 
The functions of arbitrary angles $\phi_i$ can be related to $ t_{\phi_i}$ by the formulas: 	$s_{\phi_i}=\frac{t_{\phi_i}}{\sqrt{t_{\phi_i}^2+1}},\, c_{\phi_i}=\frac{1}{\sqrt{t_{\phi_i}^2+1}}$. Furthermore, the Yukawa couplings and self-coupling of scalars 
are chosen to be smaller than $\sqrt{4\pi}$ and $4\pi$ respectively, to satisfy the limits of perturbation theory~\cite{Hung:2019jue}. The results for the dependence of $\Ga (h \rightarrow \bar{b}\, b )$ on $t_{\alpha}$ are given in Fig.\ref{fig_hbb}. The cyan line represents $\Ga (h \rightarrow \bar{b}\, b)$ and the cyan dash lines represent the upper and lower bounds from the experimental data.
In the chosen parameter space, we show that $\Ga (h \rightarrow \bar{b}\,b)$ is always within the experimental limits in the region where $t_{\beta_1} <1$ (or $v_{1}<v_{2}$). This is a very interesting result of the 331ALP model, consistent with the scalar content introduced with the hypothesis that $v_2$ generates masses for the charged leptons while $v_1$ contributes into the masses of active neutrinos after the second stage of spontaneous symmetry breaking.

\subsection{SM-like Higgs decays into two charged leptons $h \to \bar{l}l$}

Concerning the lepton sector, the Yukawa interaction for charged leptons are given by
\be
-\mathcal{L}_{Y}^{\left( l\right) } = \sum_{a=1}^{3}\sum_{b=1}^{3}g_{ab}\overline{l}_{aL}\fr{v_2 + R_2 + i I_2
}{%
	\sqrt{2}}l_{bR}+h.c. \label{LYl}
\ee

The decay rates of the processes $h \rightarrow l l$ with $l=\mu, \tau$, is
\bea
\Ga (h \rightarrow l l)& = &
\fr{g_{(h,l, l)}^2}{8\pi}m_h\left(1-\fr{4m_l^2}{m_h^2}\right)^{\fr 3{2}}= \left( \fr{v c_{\beta_1}}{v_2}\right)^2 \fr{m_l^2}{v^2} \fr{m_h}{8\pi}\left(1-\fr{4m_l^2}{m_h^2}\right)^{\fr 3{2}}\crn
& = & \left( \fr{c_{ \beta_1}}{c_{ \al}}\right)^2 \fr{m_l^2}{v^2} \fr{m_h}{8\pi}\left(1-\fr{4m_l^2}{m_h^2}\right)^{\fr 3 2}\,,
\label{Gahll}
\eea
The following constraints for the mixing angle $\beta_1$ are given by \cite{alp331}
\be
0^o \,\leq \beta_1 \leq 75^o \,,\hs \mbox{or}\,\hs 280^o \,  \leq  \beta_1 \leq 360^o \ \,.\label{conbeta1}
\ee

We use $m_{\mu}=0.105$ GeV and $m_{\tau}=1.776$ GeV with the branching ratios limited by ATLAS: $a_{h\,\bar{\mu}\,\mu}^{exp}=0.72^{+0.50}_{-0.72} \,,
a_{h\,\bar{\tau}\,\tau}^{exp}=0.93^{+0.13}_{-0.13}\,$  
(see Ref.~\cite{higgsll}).
The upper and lower bounds of the experimental data for $\Ga (h \rightarrow \bar{\mu}\, \mu)$ and $\Ga (h \rightarrow \bar{\tau}\, \tau)$ are
\bea \label{limll}
0.0   \leq \Ga (h \rightarrow \bar{\mu}\, \mu) \leq 13.5 \times 10^{-7} \, \mathrm{GeV},\,\crn \,1.66  \times 10^{-4} \leq \Ga (h \rightarrow \bar{\tau}\, \tau) \leq 2.91 \times 10^{-4}\, \mathrm{GeV}.
\eea
We also use the parameterizations from Eq.~\eq{tmixangel}.
\begin{figure}[h]
	\centering
	\begin{tabular}{c}
		\includegraphics[width=12cm]{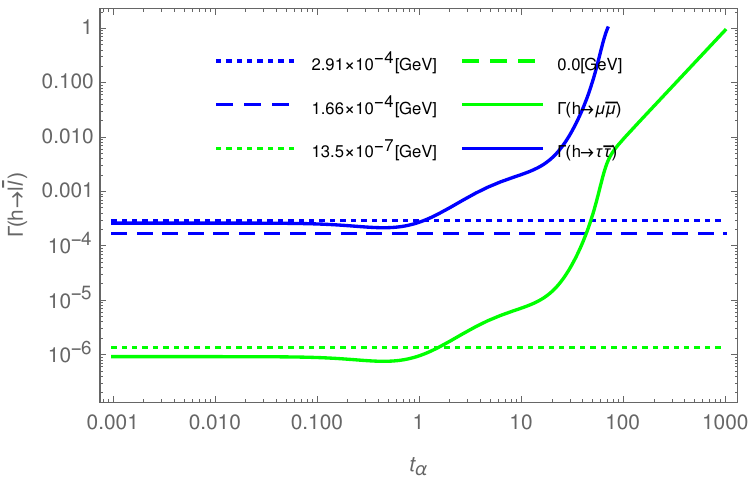}
	\end{tabular}%
	\caption{ Plots of $\Ga (h \rightarrow \bar{\mu}\, \mu)$ and $\Ga (h \rightarrow \bar{\tau}\, \tau)$ depend on $t_\alpha$.}
	\label{fig_hll}
\end{figure}
The results for the dependence of $\Ga (h \rightarrow ll)$ on $t_\al$ are given in Fig.\ref{fig_hll}. The green line represents values of $\Ga (h \rightarrow \bar{\mu}\, \mu)$ and the blue line represents values of $\Ga (h \rightarrow \bar{\tau}\, \tau)$. The parameter space region satisfies the experimental limit corresponding to $t_\alpha <1$.

\section{$PQ$ charges}\label{PQchargeSec}
Besides defining the symmetries of the model, we assume that the classical Lagrangian automatically possesses the global Peccei-Quinn symmetry.
The $PQ$ charges of multiplets were introduced in Ref.~\cite{julio2}.
Remind that the $PQ$ charge of a multiplet is an average value of all components' $PQ$ charges  in the multiplet. Under $PQ$ transformation, the fields behave as follows
\bea
f & \rightarrow & f^\prime =  e^{ i  \al \ga_5 PQ_f } f\,, \hs {\bar f} \rightarrow {\bar f}^\prime =  {\bar f} e^{ i  \al \ga_5 PQ_f } \,, \hs
\va \rightarrow \va^\prime = e^{ - i  \al  PQ_\va } \va \,.
\label{au311}\
\eea
For the convenience, the minus sign in the exponent of scalar field $\va$ is used.

For chiral fermions which are often used in the literature, there is no $\ga_5$ in the exponent
\bea
f_R &\rightarrow & f^\prime_R  =  e^{  i \al \left[PQ_f \right] } f_R\,,\hs
{\bar f}_R \rightarrow {\bar f}^\prime_R  = {\bar f}_R  e^{ - i \al  \left[PQ_f \right] }\,,\crn
f_L &\rightarrow& f^\prime_L  =  e^{ - i \al  \left[PQ_f \right]} f_L\,,\hs
{\bar f}_L \rightarrow {\bar f}^\prime_L  = {\bar f}_L  e^{  i \al \left[PQ_f \right]}\,. \label{hay1n}
\eea
Note that the combination of Eqs.\eq{au311} and \eq{hay1n} leads to the following relation
\be
PQ_f =  PQ_{\bar{f_L}} = PQ_{f_R} = - PQ_{f_L}.
\label{a161}
\ee
Within the above definition, it is obvious that the $PQ$ charge of a multiplet must follow the special rules given for chiral fermions in Eq.~\eq{au311} or in  Eq.~\eq{hay1n}.

Therefore, there are {\it two} important properties of the $PQ$ charge of a multiplet:
\ben
\item The $PQ$ charge of a left-handed multiplet is opposite in sign to the one of the corresponding right-handed multiplet;
\item The $PQ$ charge of a multiplet is opposite in sign to the one of the corresponding anti-multiplet.
\bea PQ_{\overline{f}_L} = - PQ_{f_L}\,,\hs  PQ_{\phi^*} = - PQ_{\phi}\,,\hs PQ_{(S_L)^c} = - PQ_{S_L}\,.\eea
\een
The Yukawa couplings help us to determine the $PQ$ charges. Hence, we consider the general Yukawa coupling of two fermions with a scalar
\be \mathcal{L}_{Y}^f = h  {\bar f}_L \phi \psi_R + H.c.
\lb{a162}
\ee
Under the $PQ$ transformation, it becomes
\be {\mathcal{L}^f_{Y}}^\prime  = h{\bar f}'_L \phi^\prime  \psi^\prime_R = h  {\bar f}_L  e^{  i \al  PQ_{\bar{f}_L} } \phi
e^{ - i \al  PQ_\phi}  e^{  i \al  PQ_{\psi_R} }  \psi_R + H.c.
\lb{a163}
\ee
The Lagrangian must be invariant under the $PQ$ transformations, that means ${\mathcal{L}^f_{Y}}^\prime = \mathcal{L}_{Y}^f $. As the result, the $PQ$ charge of the scalar is equal to the sum of the fermion $PQ$ charges
\be PQ_\phi =   PQ_{\bar{f}_L} + PQ_{\psi_R}\,.
\lb{a164}
\ee
This definition is consistent with result presented in the second line after Eq. (99) in Ref.~\cite{luzio} while  
the lines 9th and 10th of  Table~1 in  Ref.~\cite{julio2} do not follow this definition.

The necessary ingredients to present the $PQ$ symmetry and define the $PQ$ charges of particles are given.
Let us start defining the $PQ$ charges of fields via the Yukawa interactions of leptons with the Lagrangian as given above by Eq.~\eq{au307} in subsection~\ref{sec22}.

All terms in $\mathcal{L}_{\rm Y}^l$ in  Eq.~\eq{au307} must be invariant under the $U(1)_{PQ}$ transformation. Hence, we apply the condition of Eq.~(\ref{a164}) to all of those terms and get the following relations
\bea
PQ_{\Phi_1} &=& PQ_{\Phi_3} = - \fr 2 3 PQ_\si \,,\label{PQscalar1} \\
PQ_{\Phi_2} &=&  \fr 1 3 PQ_{\sigma}\,,\label{PQscalar2} \\
PQ_{\psi_{aL}} &=& - \fr{PQ_\sigma}{6} \,, \,
PQ_{S_{aR}} = \fr{PQ_\sigma}{2}\,,\,
PQ_{e_{aR}} = - \fr{PQ_\sigma}{2}\,. \label{PQleptons}
\eea
The combination of Eq.~\eq{PQscalar1} and Eq.~\eq{PQscalar2} leads to another intriguing relation between the $PQ$ charges of scalar triplets
\bea
\frac{PQ_{\Phi_2}}{PQ_{\Phi_1}} &=& \fr{PQ_{\Phi_2}}{PQ_{\Phi_3}} =  - \fr 1 2 \,, \label{PQratio1}\\
\fr{PQ_{\Phi_1}}{PQ_\sigma} &=& \fr{PQ_{\Phi_3}}{PQ_\sigma} = -  \frac{2}{3},
\fr{PQ_{\Phi_2}}{PQ_\sigma} = \fr{1}{3} \,.
\label{PQratio2}
\eea
The ratios $\fr{PQ_{\Phi_1}}{PQ_\sigma},\fr{PQ_{\Phi_3}}{PQ_\sigma}, \fr{PQ_{\Phi_2}}{PQ_\sigma} $ are independent from VEVs of scalar fields in the model under consideration. They are also completely different from the ratios in Eq.(38) of Ref.~\cite{julio2} where $\fr{PQ_{\Phi_2}}{PQ_{\Phi_1}} = \fr{v_1^2}{v_2^2} \neq \fr{PQ_{\Phi_2}}{PQ_{\Phi_3}} = \fr{v_3^2}{v_2^2}$. The reason of this difference is because the expression of the axion state in Eq. (19) in the Ref.~\cite{julio2} is incorrect. The correct physical state is given in Eq.~(\ref{axion1}) in subsection~\ref{CPoddSec} and in Eq.~(\ref{axion2}) in section~\ref{some}.

Now we turn back to Eq.~\eq{VH} with its last term  which is the quartic interaction $\lambda_A \sigma\, \Phi_1 \Phi_2 \Phi_3$. This term contains two important properties, namely the sum over $PQ$ charges of scalar multiplets and the general form of axion state. The $PQ$ charge of this term is defined by
\be
PQ_{\Phi_1} + PQ_{\Phi_2} + PQ_{\Phi_3} + PQ_{\si} =  - \fr{2PQ_\sigma}{3} + \fr{PQ_\sigma}{3}- \fr{2PQ_\sigma}{3} +PQ_\sigma =  0. \label{PQchargecondition}
\ee
The above condition confirms that the quartic interaction term in Eq.~(\ref{VH}), where axion and inflation appear, is really invariant under the $U(1)_{PQ}$ transformation.
It is also shown that the sum of $PQ$ charges of triplets must get the {\it opposite} sign with $PQ$ charge of singlet $\sigma$. 

Similarly to the lepton sector, in order to define the $PQ$ charges of quarks, we consider the Yukawa couplings of quarks shown in Eq.~(\ref{au306}).
All terms of $\mathcal{L}_{\rm Y}^q$ in Eq.~(\ref{au306}) must be invariant under the $U(1)_{PQ}$ transformation. So, applying the $U(1)_{PQ}$ invariance condition to all those terms in Eq.~(\ref{a164}) helps us get the following relations
\bea
PQ_{Q_{\alpha L}} &=& -   PQ_{Q_{3 L}} = \fr{PQ_\sigma}{6}\,,\crn
PQ_{u_{aR}} &=& PQ_{U_{3R}} = -  PQ_{d_{aR}} = - PQ_{D_{\alpha R}} =  \fr{PQ_\sigma}{2} \,. \label{PQquarks}
\eea
\begin{table}[h!]
	\begin{centering}
		{\renewcommand{\arraystretch}{1.5}
			\begin{tabular}{|c|c|c|c|c|c|}
				\hline
				Field & 3-3-1-1 rep & $\mathcal{B}-\mathcal{L}$  & $\mathrm{U(1)_{PQ}}$
				\tabularnewline
				\hline\hline
				$\psi_{aL}$ & $\left(\mathbf{1},\three,-\fr{1}{3},-\fr{1}{3}\right)$  & $\left(-1,-1,+1\right)^{T}$& $-\frac{PQ_\sigma}{6}$ \tabularnewline
				\hline
				$e_{aR}$ & $\left(\mathbf{1},\mathbf{1},-1,-1\right)$ & $-1$ &$ -\frac{PQ_\si}{2}$  \tabularnewline
				\hline
				$Q_{\al L}$ & $\left(\mathbf{3},\threeS,0,-\fr{1}{3}\right)$ & $\left(\fr{1}{3},\fr{1}{3},-\fr{5}{3}\right)^{T}$ & $\frac{PQ_\sigma}{6}$ \tabularnewline
				\hline
				$Q_{3L}$ & $\left(\mathbf{3},\three,\fr{1}{3},1\right)$  & $\left(\fr{1}{3},\fr{1}{3},\fr{7}{3}\right)^{T}$& $-\frac{PQ_\sigma}{6}$\tabularnewline
				\hline
				$u_{a R}$ & $\left(\mathbf{3},\mathbf{1},\fr{2}{3},\fr{1}{3}\right)$ & $\fr{1}{3}$ & $\fr{PQ_\sigma}{2}$  \tabularnewline
				\hline
				$ U_{3R}$ & $\left(\mathbf{3},\mathbf{1},\fr{2}{3},\fr{7}{3}\right)$ & $\fr{7}{3}$ & $\fr{PQ_\sigma}{2}$  \tabularnewline
				\hline
				$d_{a R}$ & $\left(\mathbf{3},\mathbf{1},-\fr{1}{3},\fr{1}{3}\right)$ & $\fr{1}{3}$ &$-\fr{PQ_\sigma}{2}$    \tabularnewline
				\hline
				$D_{\al R}$ & $\left(\mathbf{3},\mathbf{1},-\fr{1}{3},-\fr{5}{3}\right)$  & $-\fr{5}{3}$ & $-\fr{PQ_\sigma}{2}$  \tabularnewline
				\hline
				$S_{aL}$ & $\left(\mathbf{1},\mathbf{1},0,-1\right)$ & $-1$ &$-\fr{PQ_\sigma}{2}$
				\tabularnewline
				\hline
				$S_{aR}$ & $\left(\mathbf{1},\mathbf{1},0,-1\right)$ & $-1$ & $\fr{PQ_\sigma}{2}$
				\tabularnewline
				\hline
				$\Phi_{1}$ & $\left(\mathbf{1},\three,-\fr{1}{3},\fr{2}{3}\right)$  & $\left(0,0,2\right)^{T}$ &$ -\fr 2 3  PQ_\si $\tabularnewline
				\hline
				$\Phi_2$ & $\left(\mathbf{1},\three,\fr{2}{3},\fr{2}{3}\right)$ & $\left(0,0,2\right)^{T}$  & $ \fr 1 3  PQ_\si $\tabularnewline
				\hline
				$\Phi_3$ & $\left(\mathbf{1},\three,-\fr{1}{3},-\fr{4}{3}\right)$ & $\left(-2,-2,0\right)^{T}$ &$ - \fr 2 3  PQ_\si $\tabularnewline
				\hline
				$\sigma$ & $\left(\mathbf{1},\mathbf{1},0,0\right)$& $0$ & $PQ_{\sigma}$
				\tabularnewline
				\hline
		\end{tabular}}
		\par\end{centering}
	\caption{The field content, symmetry representations, and $PQ$ charges. }
	\lb{tab2}
\end{table}
The obtained $PQ$ charges of multiplets are 
summarized in the 4th column of Table \ref{tab2} where the $PQ$ charges of all fermions and scalar fields of the model can be parameterized only by $PQ_\sigma$. This result is consistent with the fact that only $\sigma$ field causes $PQ$ symmetry breaking. It is very important to study the axion-to-gluon coupling constant as well as the axion-to-photon coupling constant.
{
By imposing the $U(1)_{PQ}$ invariance to the Yukawa interactions, the $PQ$ charges of concrete (component) fields are 
 determined as shown in Table~\ref{tab2a}.
		\begin{table}[th]
			\resizebox{15cm}{!}{
				\begin{tabular}{|c|c|c|c|c|c|c|c|c|c|c|c|c|c|c|c|c|c|}
					\hline
					&  $\, u \, $ & $\, d\, $ & $U_3$ & $D_\al$ & $\, l\, $ &$ \nu$ &  $
					S$ & $\Phi^1_1 $ & $\Phi^3_1$ & $\Phi^1_3 $ & $\Phi_3^3$ & $
					\Phi_2^2$ &$\si$&$\Phi_1^2$&$\Phi_3^2$&$\Phi_2^1$&$\Phi_2^3$ \\ \hline
					$U(1)_{PQ}$ & $R$ & $-R$ & $R$ & $-R$ & $
					R$ & $-R$ &  $R$ & $-R$ & $%
					-R$ & $-R$ & $R$ & $2 R$  & $2 R$  & $0$  & $0$  & $0$  & $0$
					\\ \hline
			\end{tabular}}
			\caption{$U(1)_{PQ}$ charge assignments of the particle content of the model.
				For brevity, we use the following notation: $PQ_{\sigma} \equiv 2 R$ and  $PQ_F = PQ_{F_R} = - PQ_{F_L}$.
			}
			\label{tab2a}
		\end{table}
		
		It is seen that only neutral scalar fields ($\Phi^1_1, \Phi^3_1, \Phi^1_3, \Phi^3_3, \Phi^2_2 $) carry $PQ$ charges, while charged scalars ($\Phi_1^2,  \Phi^2_3, \Phi^1_2, \Phi^3_2 $) do not.
	}
	
	\section{Some aspects of axion\label{some}  }
	
	\subsection{Form of axion state}
	
	Being a part of $CP$-odd scalars, the physical state of axion can be defined via certain mixing angles. 
	In general, axion can be written in the form of a linear combination of imaginary parts of scalars. Moreover, the axion must be also an eigenstate of the $CP$-odd mass mixing matrix. These requirements make the searching for correct physical state of axion to be a quite complicated task.
    To make this task less complicated, a non-zero $PQ$ charge of a singlet complex scalar should be rewritten in the polar coordinates before considering the last term of Eq.~\eq{VH} which contains axion.

  With $\rho_\phi = \sqrt{\fr 1 2 (R_\phi^2 +I_\phi^2)}$ and $\tan_\phi = \fr{I_\phi}{R_\phi} $, one has
	\be \phi = \fr 1{\sqrt{2}} (R_\phi + i I_\phi) = \rho_\phi e^{i \theta_{\phi}}\,,
	\label{nov62}
	\ee
	In case $I_\phi \ll R_\phi$, Eq.~\eq{nov62} becomes
	\be  \phi \approx  \rho_\phi e^{i \fr{I_{\phi}}{R_\phi}}\,, \hs  \hs \mbox{with} \hs \rho_\phi \approx \frac{R_\phi}{\sqrt 2}\,.
	\label{nov63}
	\ee
	The complex scalar at its  VEV ($v_\phi$) is defined as follow
	\be  \phi_{v_\phi} \equiv \phi|_{R_\phi = v_\phi} =\frac{v_\phi}{\sqrt 2} e^{i \frac{I_\phi }{v_\phi}}\,.
	\label{nov64}
	\ee
	Under the $\mathrm{U(1)_X}$ transformation, the above determined singlet transforms as
	\be \phi_{v_\phi}  \rightarrow \phi_{v_\phi}^\prime = \frac{v_\phi}{\sqrt 2} e^{i X_\phi \frac{I_\phi }{v_\phi}}\,, 
	\label{nov65}
	\ee
	and $\sigma$ is unchanged because it does not carry any $X$ charge. If $U(1)_{X}$ invariance is required for the last term in Eq.~(\ref{VH}), it follows
	\begin{equation}
		\sum_{i=1}^3 I_{\Phi_i} \fr{X_{\Phi_i}}{v_{\Phi_i}} =0 \,. \label{U1Xcondition}
	\end{equation}
	Moreover, under the $U(1)_{PQ}$ transformation, one also has
	\begin{equation}
		\phi_{v_\phi}^{\prime} \to \phi_{v_\phi}^{\prime \prime} = \frac{v_\phi}{\sqrt 2} e^{i X_\phi \frac{I_\phi }{v_\phi} PQ_\phi}\,. \label{U1XU1PQ}
	\end{equation}
	
	In the special scenario where VEVs of all neutral components of $\Phi_1$ and $\Phi_3$ are {\it  unique}, these scalar fields can be rewritten as
	\begin{eqnarray}
		\Phi_1 = \langle  \Phi_1 \rangle + \Phi_1^\prime\,,\hs   \langle  \Phi_1 \rangle = \fr{v_1}{\sqrt 2}\left( \begin{array}{c}
			1 \\
			0\\
			0
		\end{array}\right)\,,\hs
		\Phi_3  =  \langle  \Phi_3 \rangle + \Phi_3^\prime\,,\hs   \langle  \Phi_3 \rangle = \fr{v_3}{\sqrt 2}\left( \begin{array}{c}
			0 \\
			0\\
			1
		\end{array}\right)\,, \label{uniqueVEVscalar}
	\end{eqnarray}
	while $\Phi_1^\prime$ and $\Phi_3^\prime$ don't have VEVs. Hence, the quartic interaction $\lambda_A \sigma \Phi_1 \Phi_2 \Phi_3$ in scalar potential for unique VEVs of scalars fields is
	\begin{eqnarray}
		\mathcal{L}_{quadrscalars}|_{_{VEV}} &  = & \lambda_A \langle  \si  \rangle \langle  \Phi_1 \rangle \langle  \Phi_2 \rangle \langle  \Phi_3 \rangle \, e^{i \left[I_\sigma \frac{PQ_\sigma}{v_\sigma}+I_{\Phi_1}\frac{X_{\Phi_1}}{v_1}\,PQ_{\Phi_1}+I_{\Phi_2}\frac{X_{\Phi_2}}{v_2}\,PQ_{\Phi_2}+I_{\Phi_3}\frac{X_{\Phi_3}}{v_3}\,PQ_{\Phi_3}\right]} \crn
		& = &
		\frac{\la_A}{4} v_\si v_1 v_2 v_3\, e^{i \left[I_\sigma \frac{PQ_\sigma}{v_\sigma}+I_{\Phi_1}\frac{X_{\Phi_1}}{v_1}\,PQ_{\Phi_1}+I_{\Phi_2}\frac{X_{\Phi_2}}{v_2}\,PQ_{\Phi_2}+I_{\Phi_3}\frac{X_{\Phi_3}}{v_3}\, PQ_{\Phi_3}\right]}\,.\label{L4scalars}
	\end{eqnarray}
	
	Finally, under the $U(1)_{PQ}$ symmetry transformation, we get the following constraint
	\begin{equation}
		I_\sigma \frac{PQ_\sigma}{v_\sigma}+I_{\Phi_1}\frac{X_{\Phi_1}}{v_1}\,PQ_{\Phi_1}+I_{\Phi_2}\frac{X_{\Phi_2}}{v_2}\,PQ_{\Phi_2}+I_{\Phi_3}\frac{X_{\Phi_3}}{v_3}\, PQ_{\Phi_3} =0 \,.\label{specialconstraint}
	\end{equation}
	Combining Eq. \eq{specialconstraint} and Eq.~\eq{modd}
		gives the following form of the axion field 
	\be
	a = \sqrt{-A}\left(\fr{  PQ_\si}{v_\si}  I_{\si}   + \sum_{i=1}^3  \fr{n_{i}}{v_i}PQ_{\phi_i} I_{\phi_i}\right) N_a^{-1}\,,
	\label{normalizedAxion}
	\ee
	where $N_a$ is a normalization factor and $ n_i, i=1,2,3$ are real numbers determined from the process of diagonalization.
	It is easily to realize that the mass term in free Lagrangian ($\fr 1 2 m^2 \, a^2$) provides mass matrix in Eq. \eq{modd}.
	It is emphasized that there are also three fields with linear combination of $\frac{I_\sigma}{\sigma}, \frac{I_i}{v_i}$ 
	and two of these are Goldstone bosons $G_Z$ and $G_{Z^\prime}$. Remind that diagonalizing process concerns on just VEVs of the model.
	On the other hand, physical state of axion must be one of four eigenstates which are determined via diagonalizing the CP-odd mass mixing matrix. This means physical state of axion must be both normalized and orthogonal with three other eigenstates. In this case, the axion  does not mix with Goldstone bosons $G_Z$ and $G_{Z^\prime}$ ~\cite{sr}.
	In the following, we will show that all coefficients $ n_i = X_{\Phi_i}, i=1,2,3.$ 
    Applying these two properties of axion, 
    the common form of axion in this model  is as below
	(see also in Ref.~\cite{sr})
	\bea a & = &\fr 1{f_{PQ}}\left( \fr{PQ_\si}{ v_\si}\, I_\si +  \sum\limits_{i=1}^3 n_
	i \fr{PQ_{\Phi_i}}{ v_i } I_{\Phi_i}\right) = \fr{  PQ_\si}{ v_\si\, f_{PQ}}\left[   I_\si + \sum\limits_{i=1}^3	\, n_i \,\fr{ v_\si }{v_i}\, \fr{PQ_{\Phi_i} }{PQ_\si} \,  I_{\Phi_i} \right]\,.
	\lb{axion2}
	\eea
	and
	$f_{PQ}$, which plays the role of normalization, is given by
	\bea
	f_{PQ} = \left[\left(\, \fr{PQ_\si}{ v_\si}\right)^2 + \sum\limits_{i=1}^3\left(n_i\,\fr{ PQ_{\Phi_i}}{ v_i}\right)^2\right]^{1/2}\,.
	\lb{a291t}
	\eea
	The normalized state of axion in Eq.~\eq{axion2} can be rewritten as
	\bea
	a    &=& \frac{\left[   I_\si + \fr 1 3 \,\fr{ PQ_{\Phi_1}}{PQ_\si}  \fr{ v_\si}{v_1} \, I_{\Phi_1}
		-  \fr 2 3 \,\fr{ PQ_{\Phi_2}}{ PQ_\si}  \fr{ v_\si}{v_2} I_{\Phi_2} + \fr 1 3 \,\fr{ PQ_{\Phi_3}}{ PQ_\si}
		\fr{ v_\si}{v_3} \, I_{\Phi_3} \right]}{\left[ 1 +  \left( \fr 1 3 \,\fr{ PQ_{\Phi_1}}{PQ_\si}  \fr{ v_\si}{v_1}\right)^2 + \left(	-  \fr 2 3 \,\fr{ PQ_{\Phi_2}}{ PQ_\si}  \fr{ v_\si}{v_2}\right)^2 +\left( \fr 1 3 \,\fr{ PQ_{\Phi_3}}{ PQ_\si}  \frac{v_\sigma}{v_3}
		\,\right)^2 \right]^{1/2}}
	\,.\label{axion3}
	\eea
    The form of axion in Eq.~\eq{axion3} depends on the ratios of $PQ$ charges of scalar triplets and $PQ$ charge of singlet $\sigma$ that means this form is independent from $PQ$ charges of scalars because all $PQ$ charges of particles are defined via $PQ_\sigma$.
	Besides, the expression of axion in Eq.~(\ref{axion1}) is rewritten as below
	\be
	a = \cos \al_2 [ I_\sigma - \sin \al_1 \tan \al_2 I_3
	- \sin \al \cos \al_1 \tan \al_2 I_2 - \cos \al  \cos \al_1 \tan \al_2 I_1]\,.
	\label{axion4}
	\ee
	Compare just numerators in Eqs. \eq{axion3} and  \eq{axion4}, one gets these below ratios
	\be
	\frac{PQ_{\Phi_2}}{PQ_{\Phi_1}} = - \fr 1 2  \,, \hs \fr{PQ_{\Phi_1}}{PQ_{\Phi_3}} = 1\,, \label{differentPQ}
	\ee
	which are consistent with the ratios determined as in Eqs.~(\ref{PQratio1}) and (\ref{PQratio2}) in Sec.~\ref{PQchargeSec}.
	
	The expression in Eq.~\eq{axion3} must be an eigenstate of possible $CP$-odd mass mixing matrix.
	In our case, within the state given in Eq.~\eq{axion1}, the normalization factor is given by
	\bea
	f_{\tilde{a}} &=& \fr{1}{\sqrt{\left[c^2 _{\alpha_2} + (s_{ \alpha_1} s_{ \alpha_2} )^2
		+ (s_{\alpha} c_{ \alpha_1} s_{ \alpha_2})^2 +
		(c_{ \alpha} c_{ \alpha_1} s_{ \alpha_2})^2\right]}}\,,\label{fa}
	\eea
	where   the absolute values in  Eq.~\eq{oddangles} take positive sign, i.e.,
	$|c_ \al| = c_ \al$ and  $|s_ {\alpha_1}| = s_{ \alpha_1} $.
	
	To finish this subsection we note that the combination of $U(1)_{PQ}$ and $SU(3)_L \times U(1)_X$ group helps to determine the general form of axion field while only $U(1)_{PQ}$ symmetry can not.
Furthermore, the axion $a$ has to be orthogonal to Goldstone bosons $G_Z$ and $G_{Z'}$.	It is emphasized that in an axion state, for scalar triplets, the $X_i$ charges are always accompanied by
	their imaginary parts $I_i$.
		
		\subsection{Axion-fermion couplings}
		\label{Fera}
		
		There are two kinds of axion-fermion couplings: the axion Yukawa couplings and the derivative (anomaly) ones that follow from $PQ$ transformations.
		Let us consider Yukawa couplings of axion with fermion.
		
		\subsubsection{Axion-fermion couplings}
		\label{yua1}
		The axion-lepton couplings appear in Eq.~\eq{au307}, where the related part
		is given by
		\be
		-\mathcal{L}_{\rm Y}^l =
		y^{e}_{ab}\,\overline{e_{aL} }\,  \Phi_2 e_{bR}   + \mathrm{H.c.}=
		y^{e}_{ab}\,\overline{e_{aL} }\,  \Phi_2 e_{bR} +
		y^{e}_{ab}\,\overline{e_{bR} }\,  (\Phi_2)^* e_{aL}
		\,.
		\lb{ju301}
		\ee
		
		From the above equation we get the following coupling of axion with leptons
		\be
		\mathcal{L}_Y^{(a \, l)}   =   \fr{i}{\sqrt{2}} y^{e}_{ab} I_{\Phi_2}\,\overline{e_{a} }\ga_5 e_b
		\,,
		\lb{ju302}
		\ee
		where \cite{julio3}
		\be
		y^{e}_{ab} = \fr{\sqrt{2} }{v_2} \rm{diag} (m_e, m_\mu, m_\tau)\de_{a b}.
		\lb{ju303}
		\ee
		Therefore interactions of axion with leptons are given by
		\be
		\mathcal{L}_Y^{(a \, l)}   =  - i \fr{ m_l }{v_2}  \left(- \sin \al \cos \al_1
		\sin \al_2 \,  a\right) \,\overline{l }\ga_5 l
		\,,\hs    \hs  (l= e, \mu,\tau)\,.
		\lb{ju305}
		\ee
         Substituting Eq.~\eq{axion1} into Eq.~\eq{ju305}, the axion-fermion Yukawa-like interactions are given as
       \bea
        \mathcal{L}_Y^{(a \, l)}   =  - i \fr{ m_l }{v_2} \fr{  PQ_\si}{ v_\si\, f_{PQ}}  \sin \al \cos \al_1		\sin \al_2 \,  \overline{l }\ga_5 l \left[   I_\si + \sum\limits_{i=1}^3	\, n_i \,\fr{ v_\si }{v_i}\, \fr{PQ_{\Phi_i} }{PQ_\si} \,  I_{\Phi_i} \right]\,. \label{LYal331}
       \eea
        From Eq.~\eq{LYal331}, the axion-leptons interaction is independent from $f_a$ and $PQ$ charges of leptons but depends on the mass mixing angles defined in $CP$ odd sector as in subsection \ref{CPoddSec}. In other words, this interaction depends on VEVs, $X$ charges and $PQ$ charges of scalar fields as well as masses of leptons. This result is consistent that the contribution of leptons to $\mathcal{L}_Y^{(a \, l)}$ should come from the VEV generating masses to leptons and the contribution of axion to $\mathcal{L}_Y^{(a \, l)}$ should come from the characteristic physical quantities of axion. It is quite different from the results of Eq.~(52) in Ref.~\cite{julio2} or Eq.~(55) of Ref.~\cite{julio3}.
				 		
		\subsubsection{Anomaly axion-fermion couplings}
		\label{yua2}
		In this work, we are interested in the anomaly couplings related to the $PQ$ transformation.  Let us start from the kinematic term of a fermion in the free Lagrangian
		\bea
		\mathcal{L}^0_f & = & \fr i 2 \bar{f}  \ga^\mu \stackrel{\leftrightarrow}{\pa_\mu}  f  =  \fr i 2 (\bar{f}  \ga^\mu \pa_\mu f - \pa_\mu \bar{f} \ga^\mu f)\,.
		\label{j261}
		\eea
		
		
		According to Ref. \cite{a2}, and using notations from \cite{gu,choi} for arbitrary fermions and scalar bosons, the $PQ$
		transformations are as follows
		\bea
		f & \rightarrow & f^\prime =  e^{ i  \left(\fr{c_f}{2 f_{pq}}\right)\ga_5 a} f\,, \hs {\bar f} \rightarrow {\bar f}^\prime =  {\bar f} e^{ i  \left(\fr{c_f}{2 f_{pq}}\right)\ga_5 a} \,, \hs
		\va \rightarrow \va^\prime = e^{ i \left(\fr{c_\va}{2 f_{pq}}\right) a} \va \,,
		\label{pqr}\\
		f_L &\rightarrow& f^\prime_L  =  e^{ - i  \left(\fr{c_f}{2 f_{pq}}\right) a} f_L\,,\hs
		{\bar f}_L \rightarrow {\bar f}^\prime_L  = {\bar f}_L  e^{  i  \left(\fr{c_f}{2 f_{pq}}\right) a}\,,\crn
		f_R &\rightarrow & f^\prime_R  =  e^{  i  \left(\fr{c_f}{2 f_{pq}}\right) a} f_R\,,\hs
		{\bar f}_R \rightarrow {\bar f}^\prime_R  = {\bar f}_R  e^{ - i  \left(\fr{c_f}{2 f_{pq}}\right) a}\,.  \label{hay1}
		\eea
		where $c_f$ and $f_{pq} \sim 10^{11} \gev$ are $PQ$ charge of the fermion and the axion decay constant related to
		the scale of the $U(1)_{PQ}$ global symmetry breaking, respectively.

		Under $PQ$ transformation as in \eq{hay1}, one gets
		anomaly axion-fermion coupling as follows
		\bea
		\mathcal{L}_ {(af)} & = &  -\fr{1}{f_{pq}}\pa_\mu  a \left[ \bar{d}\,  {\bf c}_{d} \,\ga^\mu  \ga_5 d + \bar{u}\, {\bf c}_{u}\,\ga^\mu  \ga_5 u   + \bar{U_3}\, {\bf c}_{U_3}\,  \ga^\mu  \ga_5 U_3 + \bar{D}_\al \, {\bf c}_{D_\al} \, \ga^\mu   \ga_5 D_\al + \bar{l}\,  {\bf c}_{l}\ga^\mu \, \ga_5 l   \right],\crn
		\label{s9}
		\eea
		in which the number of colors, flavor indexes and $PQ$ charges have to be counted.
		Our result here coincides with the one given in~\cite{neu}.
		Moreover, Eq.~\eq{s9} shows that derivative (anomaly) couplings
		depend on fermion $PQ$ charges (as same as EM couplings of fermions  with photon).

	\subsection{Axion - photon coupling}
	{As usual, if the mixture between scalars fields is neglected, the physical state of axion $a$ contains only one component $I_\sigma$ (as we mentioned in the last sentence of subsection \ref{CPoddSec}). In this scenario, there is an anomaly coupling of axion and a pair of photons which is arisen from} effective Lagrangian consists of the terms \cite{luzio}
	\bea
	\mathcal{L}_a  \supset \fr{a}{v_a} \fr{g_s^2 \, N}{16 \pi^2} G \tilde{G} +
	\fr{a}{v_a} \fr{e^2 E }{6 \pi^2}  F \tilde{F} + \fr{\pa^\mu a}{v_a} J_\mu^{PQ} \,,
	\lb{effecLagrangian}
	\eea
	where $\tilde{G}^{b,\mu\nu}$ is the strength field dual tensor of $G_{\mu\nu}^b$ and $g_s$ is the strong  coupling constant.
	The similarity is for values in the second term in \eq{effecLagrangian}. The $QCD$ vacuum value $v_a$ is defined as follows
	\be \langle 0 | J_\mu^{PQ} | 0 \rangle = i v_a p_\mu.
	\lb{nov182}
	\ee
	With the breaking scale of axion is given as 
	\be f_a = \frac{v_a}{2 N}\,,
	\lb{nov183}
	\ee
	and this makes Eq.~\eq{effecLagrangian} to become
	\bea
	\mathcal{L}_a  \supset \fr{a}{f_a} \fr{g_s^2 \, }{32 \pi^2} G \tilde{G} +
	\fr{1}{4} g_{a\, \ga}^0  a F \tilde{F} + \fr{\pa^\mu a}{v_a} J_\mu^{PQ}\,,
	\lb{effecLagran1}
	\eea
	where
	\be g_{a\, \ga}^0 = \fr{\al}{2 \, \pi  \, f_a}\, \fr E N \,.
	\lb{anomalycoupling}
	\ee
	The axion-photon {anomaly} coupling in Eq.~\eq{anomalycoupling} {appears only when the global symmetry $U(1)_{PQ}$ is spontaneously broken by $v_\sigma$ at energy scale of $10^{12} \gev$. This anomaly coupling} is very attractive because it plays an important role for experiments searching for axion such as CERN Axion Solar Telescope (CAST), etc, via Primakoff effect \cite{agu,s5,s6,s7}.
	In the model under consideration, the QCD anomaly coefficient is defined as~\cite{luzio}
	\be
	\label{nov4}
	N=\sum_\mathcal{Q} N_\mathcal{Q} = \sum_\mathcal{Q} PQ({\mathcal{Q})}\,
	n_C (\mathcal{Q})\,  n_I(\mathcal{Q})\,  T(\cal{C}_{\mathcal{Q}}) \,,
	\ee
	in which
	\begin{enumerate}
		\item $ n_C (\mathcal{Q})$ and $n_I(\mathcal{Q})$ denote the dimension of the color and weak isospin representations, respectively, for example  ($n( {\bf  T) }= 3$ for triplet and $n ({\bf S})	= 1$ for singlet).
		\item $T(\cal{C}_\mathcal{Q})$ is the colour Dynkin index, for example
		$T(3)=1/2$.
	\end{enumerate}
	
	Taking into account that $PQ_{ q_L} = - PQ_{q_R}$, we see that, in the model under consideration, quantity $N$ reads
	\be N_{331} = N(Q_{\al L}) + N(Q_{3 L}) - N(u_{a R}) - N(d_{a R}) - N(U_{3 R}) - N(D_{\al  R})\,,
	\lb{nov186}
	\ee
	where
	\bea
	N(Q_{\al L}) & = & \fr{PQ_\si}6 . 3 . 2.  \fr 1 2 =  \fr{PQ_\si}2\,,\crn
	N(Q_{3 L})  & = &  - \fr{PQ_\si}6 . 3 . 1.  \fr 1 2 = - \fr{PQ_\si}4\,,\crn
	N(u_{a R}) & = &   \fr{PQ_\si}2 . 3 . 3.  \fr 1 2 =   9.  \fr{PQ_\si}4\,,\lb{nov187}\\
	N(d_{a R}) & = &  - \fr{PQ_\si}2 . 3 . 3.  \fr 1 2 =  -  9.  \fr{PQ_\si}4\,,\crn
	N(U_{3 R}) & = &   \fr{PQ_\si}2 . 3 . 1.  \fr 1 2 =   3 \fr{PQ_\si}4\,,\crn
	N(D_{\al  R}) & = & -  \fr{PQ_\si}2 . 3 . 2.  \fr 1 2 =  -  3\fr{PQ_\si}2\,\,.\nn
	\eea
	Substituting Eq.~\eq{nov187} into Eq.~\eq{nov186}, one gets
	\be
	N_{331} = PQ_\si  \,.
	\lb{nov188}
	\ee
	
	Now we turn to the electromagnetic $[U(1)_Q]^2\times U(1)_{PQ}$ anomaly coefficient
	\be
	E  = \sum\limits_{i= charged}\left( PQ_{f_{iL}} - PQ_{f_{i R}}\right)(Q_{f_i})^2\,.
	\lb{nov189}
	\ee	
	Taking $PQ$ charges from Table \ref{tab2} and the number of quark colors  $n_c $ to be three, one gets
	\bea	E_{331}	&=& n_c \sum\limits_{u_i}^{i=1,2,3} 2 PQ_{u_{iL}} Q_{u_{i}}^2
	+ n_c\sum\limits_{d_i}^{i=1,2,3} 2 PQ_{d_{iL}} Q_{d_{i}}^2
	+n_c \sum\limits_{D_{\alpha L}}^{\alpha=1,2} 2 PQ_{D_{\alpha L}} Q_{D_{\alpha}}^2 \notag\\
    &&	+ n_c 2 PQ_{U_{3L}}  Q_{U_{3 L}}^2
	+\sum\limits_{e_i}^{i=1,2,3} 2 PQ_{e_{iL}} Q_{e_{i}}^2 \notag\\
	&=& 6\,  n_c \, PQ_{u_{a L}} \left( \frac{2}{3} \right)^2  + 2\,  n_c \, PQ_{U_{3L}} \left( \frac{2}{3} \right)^2	+6\,  n_c \,   PQ_{d_{a L}} \left( -\frac{1}{3} \right)^2  \notag\\
    &&+ 4\,  n_c PQ_{D_{\alpha L}} \left( -\frac{1}{3} \right)^2  + 6 PQ_{e_{aL}} (-1)^2\notag\\
	&=&  18 \left(-\frac{PQ_{\sigma}}{2}\right)\frac{4}{9} + 6 \left(-\frac{PQ_{\sigma}}{2}\right)\frac{4}{9}+ 18 \frac{PQ_{\sigma}}{2}\frac{1}{9}+ 12  \frac{PQ_{\sigma}}{2}\frac{1}{9} +6\left(\frac{PQ_{\sigma}}{2}\right) \notag\\
	&=& 
	- \fr 4 3 PQ_\si \,.
	\label{effecLagrangian0}
	\eea
	Then, in the model under consideration, one gets
	\be
	\fr{E_{331}}{N_{331}} = - \fr{4}{3}\,. \label{ENratio}
	\ee
	Performing current algebra, we get the axion mass given by~\cite{sr}
	\be
	m_a= \frac{(m_u m_d)^{\fr{1}{2}}}{m_u + m_d} \fr{m_\pi f_\pi}{v_\sigma} \simeq 5.7 \left( \frac{10^{12}\gev}{v_\sigma}\right) {\rm \mu eV}\,.\label{massaxion}
	\ee
	and
	\be
	g_{a\ga} =\fr{\al}{2 \pi f_a} \left(\fr E N - \fr{2}{3}\fr{4+\frac{m_u}{m_d} + \fr{m_u}{m_s} }{1+\fr{m_u}{m_d}  +\fr{m_u}{m_s}}\right) = \fr{\al}{2 \pi f_a} \left(\fr E N  - 1.92 \right) \,.
	\label{gagacite}
	\ee
	Substituting Eq.~\eq{ENratio} into Eq.~\eq{gagacite}, we get the axion-photon coupling as follows
	\be
	g_{a\ga}^{331}  = \fr{\al}{2 \pi f_a} \left(- \fr 4 3  - 1.92 \right) \,.
	\label{gaga331}
	\ee
	This result is consistent with the one given in Ref.~\cite{julio2}
	and its absolute value is larger than the one in the  DFSZ-II model~\cite{DFSZ1,DFSZ2} where the ratio $\frac{E}{N} = + \frac{2}{3}$. 

\section{Axion decay into two photons\lb{Agaga}}
Decay of an axion into two photons is studied via the following process
\bea
a(k) \rightarrow \gamma (p) + \gamma (q)\,.
\label{agaga}
\eea
{In our scenario, the mixtures of all scalar fields are considered, axion $a$ is presented via four components with three mixing angles as in Eq.~\eq{axion1}. The imaginary component $I_\sigma$ of singlet $\sigma$ with VEV $v_\sigma$ breaking $U(1)_{PQ}$ global symmetry generates the anomaly coupling which is arisen from the below effective Lagrangian
\bea
\mathcal{L}_{g_{a\ga}} = g_{a\ga}\,aF\tilde{F}\,. \label{agaga1}
\eea
Moreover, three other imaginary components $I_i$, $(i=1,2,3)$ of triplets $\Phi_i$ with their VEVs $v_i$ breaking the local symmetries $SU(3)_L \times U(1)_X \to SU(2)_L \times U(1)_Y \to U(1)_Q$, generate the interactions of axion with a pair of photons which are arisen from 
the kinetic terms of the scalar fields} in Eq.~\eq{kineticGB} where the covariant derivative is defined as
	\bea
	D_\mu & = & \pa^\mu + i g W_\mu^a T_a  + i g_X B_\mu X \,,
	\lb{covariantD}
	\eea
	with $T_a,(a=\overline{1,8}) $ being $SU(3)_L$ generators. The covariant derivatives of triplets are
	\bea
	D_\mu \phi_i & = & \left(\pa^\mu + i \fr g 2 W_\mu^a \la_a + i g_X B_\mu X \right) \phi_i \equiv  \left[\pa^\mu + i P_\mu^{CC} + i P_\mu^{NC}\right]\phi_i\,,
	\lb{j243}
	\eea
	where the charged and neutral currents are given as
	\bea P_\mu^{CC} & = & \fr g 2
	\sum_{a=1}^{8}
W_\mu^a \la_a\,,\\
P_\mu^{NC} & = &  \fr g 2\left( \sum_{a= 3,8}  W_\mu^a \la_a + 2\frac{g_X}{g}\,  X\, B_\mu   \right)\,.
\lb{j244}
\eea
{The new kind of interaction of $a \to \ga \ga$ is available only when the $CP$-odd mass mixing matrix is exactly diagonalized as presented in subsection \ref{CPoddSec} where physical state of axion with four components is pointed out. To determine this kind of interaction,} we make a shift of scalar fields by their VEVs as follows
\be \phi_{i} \rightarrow \langle \phi_{i}  \rangle + \phi_{i} \,,
\lb{o221}
\ee
and substitute {those expanded scalar fields} into Eq.~\eq{kineticGB}. It {is shown}  that the neutral interactions between scalars and gauge bosons are
\bea
\mathcal{L}_{neutral} & = & 
{\sum_{i=1,2,3} \pa^\mu \Phi_i \pa_\mu \Phi_i} + i \sum_{i=1,2,3}\left[\, \langle \Phi_i \rangle^\dag
P_\mu^{NC} \pa^\mu  \Phi_i - \pa_\mu \phi_i^\dag P^{NC, \mu}  \langle \Phi_i \rangle \right]
\crn
&&+ i \left[ \Phi_i^\dag P^{NC}_{\mu} \pa^\mu \Phi_i - \pa_\mu \Phi_i^\dag  P^{NC, \mu}\Phi_i
\right] + \Phi_i^{\, \dag}  P_\mu^{NC}  P^{NC, \mu}\Phi_i\,
+  \langle \Phi_i \rangle^\dag   P_\mu^{NC}  P^{NC, \mu}  \langle \Phi_i \rangle \crn
&&+   \langle \Phi_i \rangle^\dag   P_\mu^{NC}  P^{NC, \mu}\Phi_i +  \Phi_i^\dag  P_\mu^{NC}  P^{NC, \mu}  \langle \Phi_i \rangle,.
\lb{Lneutral}
\eea
The third term in the second line of \eq{Lneutral} provides masses of gauge bosons given in Eqs. \eq{mWXY} and \eq{mAZZp}, while the terms in the third line provide the coupling of two gauge bosons and one scalar field. Because the electromagnetic field $A_\mu$ is a
neutral gauge boson which belongs to $P_\mu^{NC}$, the interaction of interest $a \to \ga \ga$ can be studied by considering the terms in the third line of Eq.~\eq{Lneutral}.
{Using expansions in Eq.~\eq{VEVtriplet}, the last term in Eq.~\eq{Lneutral} gives
\bea
\mathcal{L}_{a \ga \ga} \propto  i \fr{g^2}4  A_\mu A^\mu \sum_{i=1}^3 v_i I_i  \,
\label{agaga2}
\eea
in which, $I_i$ can be determined from Eq.~\eq{physodd}. It is easy to see that these imaginary components contain axion $a$. Ultimately, the interaction in Eq.~\eq{agaga2} is rewritten as
\bea
 \mathcal{L}_{C_{a\ga}} = i\,C_{a\ga}\,a A_\mu A^\mu\,, \label{agaga3}
\eea
with the coupling constant includes all VEVs of triplet scalars and is determined as}
\bea
C_{a\ga}= \frac{g^2}{128}\left(10+11c_{2W} \right) t_W^2\,s_{\alpha_2}\left( v_1 c_{\alpha}c_{\alpha_1} + 4 v_2 s_{\alpha}c_{\alpha_1} +v_3 s_{\alpha_1} \right)\,. \label{Cagaga}
\eea
{Substituting Eq.~(\ref{oddangles}) into Eq.~(\ref{Cagaga}), the coupling constant $C_{a\gamma}$ is rewritten as
\bea
C_{a\ga}= \frac{3g^2 v_\sigma}{64}\left(10+11c_{2W} \right) t_W^2\,s_{\alpha_2}t_{\alpha_2}\,.
\eea}
{And total Lagrangian of $a \to \gamma \gamma$ process is
\bea
\mathcal{L}_{a\to\gamma\gamma} = \mathcal{L}_{g_{a\gamma}} + \mathcal{L}_{C_{a\gamma}} \,. \label{Lagaga}
\eea}
{These below useful formulas }
\be    \ep_{\mu \nu \al \beta}\ep^{\la \rho \si \tau}=-
\left| \begin{array}{cccc}
	\de_\mu^\la &  \de_\nu^\la & \de_\al^\la &  \de_\beta^\la \\
	\de_\mu^\rho &  \de_\nu^\rho  &  \de_\al^\rho  &  \de_\beta^\rho  \\
	\de_\mu^\si &  \de_\nu^\si  &   \de_\al^\si &   \de_\beta^\si \\
	\de_\mu^\tau &   \de_\nu^\tau &  \de_\al^\tau &   \de_\beta^\tau \\
\end{array}  \right| , \hs
\ep_{\mu \nu \al \beta}\ep^{\la \rho \si \beta}=
\left| \begin{array}{ccc}
	\de_\mu^\la &  \de_\nu^\la & \de_\al^\la   \\
	\de_\mu^\rho &  \de_\nu^\rho  &  \de_\al^\rho    \\
	\de_\mu^\si &  \de_\nu^\si  &   \de_\al^\si    \\
\end{array}  \right|,\lb{ma31}
\ee
\be   \ep_{\mu \nu \al \beta}\ep^{\la \rho \al \beta}=
-2(\de_\mu^\la\de_\nu^\rho-\de_\mu^\rho\de_\nu^\la),\
\ep_{\mu \nu \al \beta}\ep^{\la \nu \al
	\beta}=-6\de_\mu^\la,\
\ep_{\mu \nu \al \beta}\ep^{\mu \nu \al \beta}=-24\,, \lb{ma32}
\ee
{are used to calculate the decay amplitude of the above triple-couplings of axion-photon-photon have the decay amplitude as
\bea
\mathcal{M}_{g_{a\gamma}} &=& i 2 \, g_{a \gamma} \,  \epsilon^{\mu \nu \alpha \beta}\,  p_\alpha \, q_\beta \ep(p,\la)_\mu \ep(q,\si)_\nu\,,\\
\mathcal{M}_{C_{a\gamma}} &=&
- 2\,  \, C_{a \gamma} \,\,  \eta^{\mu \nu } \ep(p,\la)_\mu \ep(q,\si)_\nu\,.
\lb{MCagg}
\eea}
Hence, the squared amplitudes are
\bea
|\mathcal{M}_{C_{a\gamma}}|^2  =   16 C^2_{a \gamma}\,,\hs
|\mathcal{M}_{g_{a\gamma}}|^2  =  4 g^2_{a \gamma} m_a^4 
\,.
\lb{ma51}
\eea
{In general, the decay width of the process $a \to \ga \ga$ is
\be
\Ga_{(a  \rightarrow \ga  \ga)}  = \fr{|\mathcal{M}_{(a \\rightarrow \ga  \ga)}|^2}{64 \pi m_a}\,.\label{generaldecaywidth}
\ee
Eq.~(\ref{generaldecaywidth}) is used to calculate the decay width of two contributions to the process $a \to \ga \ga$ as
\bea
\Ga_{C_{a\gamma}}
= \frac{C_{a\ga}^2}{4\pi m_a}\,,\hs
\Ga_{g_{a\gamma}} 
=\frac{g_{a\ga}^2 m_a^3}{16 \pi} 
\,. \label{Gaagaga}
\eea
}
{The ratio between these two contributions of decay width in Eq.~\eq{Gaagaga} helps to estimate the value of new contribution in comparison with the well-known contribution. The considered ratio is
\bea
\frac{\Gamma_{C_{a\gamma}}}{\Gamma_{g_{a\gamma}}} = \frac{4 C^2_{a\gamma}}{g^2_{a\gamma} \, m_a^4}\,, \label{ratioCg}
\eea}
with mass of axion is given in Eq.~\eq{massaxion} and couplings $g_{a\gamma}, C_{a\gamma}$ are given in Eqs.~\eq{gaga331}, \eq{Cagaga}, respectively. In scenario that $\Gamma_{C_{a\gamma}}=\Gamma_{g_{a\gamma}}$, constraint for mass of axion is
\be m_a^2 = \frac{2|C_{a\gamma}|}{|g_{a\gamma}|}\,.\label{constraintmassaxion}
\ee
Combine with Eq.~(\ref{massaxion}), it is shown that $|C_{a\gamma}| \propto 10^{-12} |g_{a\gamma}|$. In this case, this new contribution has a bit significance for light axion searching.

{From another view point with heavier axion gets masses from some $\mu$eV to keV, we assume that $v_1, v_2$ is at EW scale, $v_3 \sim 10^5$ GeV, $v_\sigma \sim 10^{12}$ GeV, $f_a = 10^{12}$ GeV and coupling constant of $SU(3)_L$ is $g =10^{-7} \div 10^{-6}$ while values of Weinberg angle from experiment are $\sin^2\theta^{eff}_{W}=0.23142 \pm 0.00073$
 \cite{WeinbergAngle}, numerical analysis for the ratio in Eq.~(\ref{ratioCg}) is studied and presented by various colorful dots in Fig.~\ref{GaggRatio}.
\begin{figure}
	[h!]
	\centering
	\begin{tabular}{c}
		  \includegraphics[width=12cm]{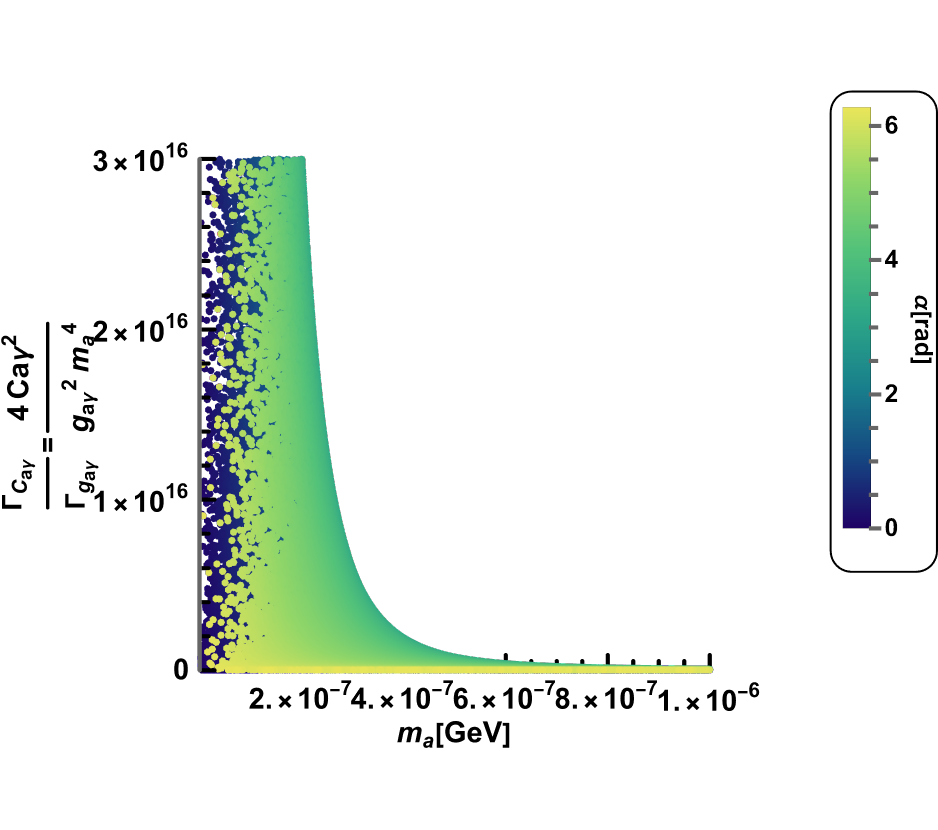}
	\end{tabular}%
	\caption{The ratio between two types of contributions to decay width of $a \to  \gamma \gamma$ depends on the mass mixing angle $\alpha$ given by VEVs at electroweak scale.}
	\label{GaggRatio}
\end{figure}
It is easily to see that as the axion mass decreases from keV to a few hundred eV, the ratio between the two contributions increases very rapidly. When $m_a=2.10^{-7}$ GeV, this ratio reaches a value of about $10^{16}$ for most values of the mixing angle $\alpha$. When mass of axion is smaller than $10^{-7}$ GeV, values of the ratio $\frac{\Gamma_{C_{a\gamma}}}{\Gamma_{g_{a\gamma}}}$ strongly depend on the mass mixing angle $\alpha$. With $\alpha <1$ rad, the ratios are almost presented by dark teal dots and get values from zero to $10^{16}$ and above while $|C_{a\gamma}| \propto 10^{-7} |g_{a\gamma}|$. But when $\alpha$ gets maximal, the ratios are presented by yellow dots and their own values are near to zero. \\
Numerical analysis shows that the new interaction of triple-coupling axion-photon-photon arisen from kinetic terms of scalars, can strongly contribute to width decay of the process when considering to axion with mass at keV scale.}
{And note again that the new interaction of the triple-coupling axion-photon-photon
  is a new result which initial appears in theory relating to axion.}

\section{Conclusions and outlooks}
\lb{sec6}
In this paper, the realization of the $PQ$ mechanism is described within the 3-3-1 model with additional $\mathcal{B-L}$ symmetry. By applying two properties of the $PQ$ charge of a multiplet that are proposed in Sec.~\ref{PQchargeSec} and assuming that the $PQ$ symmetry of Yukawa interactions is preserved, it is demonstrated that $PQ$ charges of all fermions and scalar fields in the model can be expressed in terms of the $PQ$ charge of the singlet complex scalar field which causes the breaking of $U(1)_{PQ}$ symmetry. In particular, all $PQ$ charges of scalar triplets are related to that one of the singlet $\si$, namely $PQ_{\Phi_1}= PQ_{\Phi_3} = -2  PQ_{\Phi_2} = - \fr 2 3 PQ_\si$.  It is shown that the charged scalar fields do not carry $PQ$ charged and there are no triple couplings of axion with a pair of charged gauge bosons such as $W^- W^+, Y^- Y^+, X^0 X^{0*}$. Moreover, $PQ$ charges of each components in triplets of the model are also defined. All of these $PQ$ charges are independent from the others except $PQ_\sigma$. Hence, the ratios of $PQ$ charges of the scalar fields are independent from their VEVs.

The diagonalization of the $CP$ even scalar sector points out the SM-like Higgs boson of the model under consideration. That allowed us to study SM-like Higgs decays into two down-type quarks $h \to \bar{b} b$ and SM-like Higgs decays into two charged leptons $h \to \bar{\mu} \mu$, $h \to \bar{\tau} \tau$ and the dependency of decay widths on the mixing angle $\alpha$. These results are consistent with the existing experimental bounds.

Besides, the diagonalization of the $CP$ odd scalar sector is explicitly performed to show the physical state of axion which is consistent with the one defined by requiring the preservation of $U(1)_{PQ}$ global symmetry as well as the invariance of $SU(3)_L \times U(1)_X$ transformation for all imaginary parts of scalars. 
The expression of axion state {is in a linear combination of all imaginary parts of scalar fields. These imaginary components are together with their respective $X$ charges while scalars' VEVs lie in the denominators of axion expression}.
 This form is completely  different from that in Eq.~(39) of Ref. \cite{julio2} where {imaginary parts of scalars $I_i$, ($i=1,2,3$) are linearly combined with their corresponding VEVs.} 

Moreover, the derivative couplings of axion with fermions depend on $PQ$ charges of the latter, but the axion-fermion Yukawa-like couplings are not. The mass of the axion field and its interactions with photons are derived. The decay	of axion into a pair of photons ($a \to \ga \ga$) consists of two parts. The first one is related to the anomaly coupling $g_{a \ga}$ and the second one $C_{a \ga}$ arises { from kinetic terms of scalars. Their contribution to decay width of process $a\to\gamma\gamma$ are numerically studied. The new interaction can help to search for axion with mass around hundred keV.} 

\section*{Acknowledgements}
This research has received funding from  
National Foundation for Science and Technology Development (NAFOSTED) under grant number 103.01-2023.45.


\appendix



\end{document}